%% file: full.tex
\documentclass[twocolumn]{autart}     % Enable this line and disable the 
\usepackage{graphicx}
\usepackage{subcaption}
\usepackage{amsmath}
\usepackage{amssymb}
\usepackage{accents}
\usepackage{cases}
\usepackage{tikz,pgfplots}
\pgfplotsset{compat=newest}
\usepackage{xcolor}
\usepackage{caption}
\usepackage{setspace}
\usepackage{booktabs}
\usepackage{array}
\usepackage{cite}
\usepackage{algorithm}
\usepackage{algorithmic}
\usepackage{natbib}
\usetikzlibrary{patterns}
\usetikzlibrary{graphs}
\usetikzlibrary{graphs.standard}

\newtheorem{proposition}{Proposition}
\newtheorem{theorem}{Theorem}
\newtheorem{definition}{Definition}
\newtheorem{lemma}{Lemma}
\newtheorem{corollary}{Corollary}
\newtheorem{remark}{Remark}
\newtheorem{assumption}{Assumption}

\newcommand{\revise}[1]{\textcolor{black}{#1}}

\DeclareMathOperator{\1}{\textbf{1}} 
\DeclareMathOperator*{\col}{col}
\DeclareMathOperator*{\adj}{adj}

\makeatletter
\pgfmathdeclarefunction{alpha}{1}{%
	\pgfmathint@{#1}%
	\edef\pgfmathresult{\pgffor@alpha{\pgfmathresult}}%
}

\begin{document}

\begin{frontmatter}
	
	\title{Distributed Parameter Estimation with Gaussian Observation Noises in Time-varying Digraphs \thanksref{footnoteinfo}}               % Title, preferably not more 

	\thanks[footnoteinfo]{This work was supported in the part by JSPS under Grants-in-Aid for Scientific Research Grant No. 22H01508 and 21F40376. It was also supported by the Swiss National Science Foundation supported this work through NCCR Automation under Grant agreement 51NF40\_180545.}

\author[ETH]{Jiaqi Yan}\ead{jqyan@buaa.edu.cn},    % Add the 
\author[TITECH]{Hideaki Ishii}\ead{hideaki\_ishii@ipc.i.u-tokyo.ac.jp}               % e-mail address 
% (ead) as shown
\address[ETH]{Jiaqi Yan is with the School of Automation Science and Electrical Engineering, Beihang University, Beijing 100191, P. R. China}
\address[TITECH]{Department of Information Physics and Computing,
	The University of Tokyo, Tokyo 113-8656, Japan}

	                             % keyword list or with the 
	% help of the Automatica 
	% keyword wizard

	\begin{abstract}                          % Abstract of not more than 200 words.
In this paper, we consider the problem of distributed parameter estimation in sensor networks. Each sensor makes successive observations of an unknown $d$-dimensional parameter, which might be subject to Gaussian random noises. The sensors aim to infer the true value of the unknown parameter by cooperating with each other. 
To this end, we first generalize the so-called dynamic regressor extension and mixing (DREM) algorithm to stochastic systems, with which the problem of estimating a $d$-dimensional vector parameter is transformed to that of $d$ scalar ones: one for each of the unknown parameters. 
For each of the scalar problem, both combine-then-adapt (CTA) and adapt-then-combine (ATC) diffusion-based estimation algorithms are given, where each sensor performs a combination step to fuse the local estimates in its in-neighborhood, alongside an adaptation step to process its streaming observations. Under weak conditions on network topology and excitation of regressors, we show that the proposed estimators guarantee that each sensor infers the true parameter, even if any individual of them cannot by itself.  
Specifically, it is required that the union of topologies over an interval with fixed length is strongly connected. Moreover, the sensors must collectively satisfy a cooperative persistent excitation (PE) condition, which relaxes the traditional PE condition. 
%Without stringent requirements such as statistical independencyand stationarity, the cooperative PE condition can be easily met even in stochastic systems with feedback.
Numerical examples are finally provided to illustrate the established results.
	\end{abstract}
	
\end{frontmatter}

%%%%%%%%%%%%%%%%%%%%%%%%%%%%%%%%%%%%%%%%%%%%%%%%%%%%%%%%%%%%%%%%%%%%%%%
\section{Introduction}
As a fundamental problem appearing in various applications such as signal processing, system identification and adaptive control, parameter estimation has been extensively studied in the literature for decades (see, for example, \cite{goodwin2014adaptive,xie2020convergence,schizas2009distributed,chen2013distributed,cattivelli2008diffusion,lopes2008diffusion}). In this problem, each sensor observes (partial) information of a system with an unknown (vector) parameter, and attempts to infer the true parameter through a stream of observations. In a centralized setting where all measurements are collected in a center, it is well known that consistent estimation is possible when its regressor meets certain excitation conditions (\hspace{1pt}\cite{goodwin2014adaptive}). Moreover, a \textit{persistent excitation} (PE) condition, which requires that the input signals are sufficiently rich such that all modes of the system can be excited, is usually needed to achieve exponential convergence in estimation.

By contrast, in a distributed setting, a network of multiple sensors estimates the state cooperatively by interacting with each other. 
Then, the PE condition may not necessarily hold for each individual sensor. As such, the local information available at each sensor's side is not enough to reconstruct the unknown parameter in full. To solve this problem, researchers have leveraged the communication among sensors and introduced distributed algorithms into the design of estimation schemes (\hspace{1pt}\cite{schizas2007distributed,cattivelli2009diffusion,kar2012distributed,yan2022distributed,zhang2012distributed,matveev2021diffusion,yan2021resilient,fang2020graph}). In these works, sensors update their local estimates by integrating recent sensing information and merging the estimates of neighboring sensors. As a result, weaker excitation conditions have been proposed under which the estimation task can be cooperatively fulfilled by the entire sensor network. For example, in \cite{chen2013distributed}, by using consensus
theory, the authors have proposed a cooperative PE condition, with which sensors collectively satisfy the PE condition even if no individual sensor can do so. 
%In the recent work \cite{matveev2021diffusion}, it has also been shown that a single sensor satisfying the PE condition can lead all the others to converge to the true parameter provided that this sensor is sufficiently connected by the others. 

Different from the above results primarily focusing on deterministic systems, another group of works further extends its scope to include scenarios where sensors' measurements could be affected by Gaussian white noises. Most of the existing works require that the regressors are generated under certain statistical independence or stationarity conditions. For example, in \cite{abdolee2014diffusion, takahashi2010diffusion,yu2019robust}, performance of the proposed estimators has been studied for networks where the regessors and measurement noises are independently
and identically distributed (i.i.d.) in both time and space. Similarly, \cite{tu2012diffusion} has developed distributed estimators provided that the regressors are temporally white and spatially independent. Despite the elegant results in these works, the independence and stationarity assumptions are easily violated when the regressors are generated from feedback systems. In this regard, 
\cite{xie2020convergence} and \cite{xie2018analysis} have relaxed this assumption and proposed a cooperative stochastic information condition. By developing adaptive filters based on diffusion algorithms, they have shown that the local estimate of each sensor converges to a small neighborhood of the true parameter. Yet, it is also noted that the cooperative stochastic information condition therein is difficult to verify, since it is developed based on a conditional expectation on all historical information.

Inspired by these works, this paper also investigates the problem of distributed parameter estimation under Gaussian observation noises. 
We propose distributed estimators utilizing diffusion strategies. The diffusion-based estimators usually involve a two-step process for each sensor: \textit{combination}, which fuses the neighboring estimates through a linear combination, and \textit{adaptation}, which updates the local estimate by incorporating new observations (\cite{lopes2008diffusion,xie2018analysis,gan2022convergence,tu2012diffusion,nosrati2015adaptive,takahashi2010diffusion}).
Depending on the order
of these two steps, this paper presents both combine-then-adapt (CTA) and adapt-then-combine (ATC) algorithms, where an average consensus algorithm is applied in the combination step, and a least-square approach is applied in the adaptation step, utilizing the dynamic regressor extension and mixing (DREM) algorithm.

The DREM algorithm was first introduced in \cite{aranovskiy2017performance}. The main feature of it is
to transform the problem of estimating a vector parameter into a set
of scalar ones, which allows each unknown parameter to be independently inferred. Since then, the analysis of DREM has emerged in various scenarios (\cite{aranovskiy2017performance,ortega2020new,pyrkin2019adaptive,yi2022conditions,bobtsov2022generation}). For instance, \cite{matveev2021diffusion} have extended DREM to a distributed framework, where a network of deterministic subsystems is shown to collectively identify the unknown parameter. Particularly, a ``root sensor'' that verifies the PE condition is required. With a spanning tree rooted at this sensor, it leads all the sensors to converge to the true parameter. Additionally, in \cite{wang2019robust}, DREM's behavior is explored in the presence of deterministic and bounded noises. Utilizing the input-to-state stability (ISS) theory, the study demonstrates the boundedness of estimation errors. In contrast to these works, our paper tackles the challenge of dealing with stochastic and unbounded Gaussian noises within a distributed network, which prevents direct application of the ISS theory. Consequently, we resort to the stochastic approximation method and probability limit theory for our analysis. To be specific, our main contributions are summarized as below:

	(1) Our paper introduces a novel framework for analyzing estimation performance. Specifically, we decompose a stochastic Lyapunov function into two parts, which respectively correspond to the average and variance of local estimation errors. Using probability limit theory, we show that the first part decreases solely in the combination step when local estimates are mixed by the consensus algorithm. In contrast, the second part decreases solely in the adaptation step when certain sensors are sufficiently excited. This decoupled structure enables us to independently design and analyze the combination and adaptation steps. Consequently, although the adaptation step is designed by using DREM in this paper, it can be substituted with other models. To our knowledge, this is the first time that this decoupling result has been proposed in this field.

(2) The estimation performance hinges on two factors: the richness of regressors (excitation condition) and the level of information exchange within the network (topology condition). Our performance is guaranteed under weak excitation and topology conditions. To be specific, for the excitation condition, the sensors should collectively satisfy the cooperative PE condition, which relaxes the conventional PE, without requiring statistical independency or stationarity. Moreover, the topology condition only requires that the union of digraphs over a fixed-length interval is strongly connected. These requirements are notably less stringent than those found in existing works, including \cite{abdolee2014diffusion,xie2018analysis,matveev2021diffusion,yan2023distributed} in Table~\ref{tab:compare}, as well as others like \cite{chen2013distributed,tu2012diffusion,xie2020convergence}. 

(3) We show that every sensor can infer the true parameter by cooperating with others, even when individual sensors are not persistently excited, and the network may experience disconnections. It is important to differentiate our algorithm from some prior works, such as \cite{abdolee2014diffusion, xie2018analysis, xie2015stability}, where only mean-square or $L_p$ stability of the estimation error is obtained. This suggests that the local estimate of each sensor converges to a neighborhood of the true parameter. In contrast, our algorithm provides a more precise result by guaranteeing that the local estimates strictly converge to the true parameter in a mean square sense.

\begin{table}[t!]\small
	\centering
	\begin{tabular}{p{1.7cm}p{1.6cm}p{1.7cm}p{2.0cm}}
		\toprule[1.2pt]
		Algorithm  &Excitation condition &Topology condition &Convergence of estimation error\\
		\midrule[1.2pt]
		\cite{abdolee2014diffusion} &Regressors are i.i.d.  &Fixed and connected undirected graph &Mean-square stability
		\\
		\midrule
		\cite{xie2018analysis} &Cooperative stochastic information condition &Fixed and connected undirected graph  &$L_p$ stability\\
		\midrule
		\cite{matveev2021diffusion} &PE holds at certain sensor $s$	&Contain a spanning tree rooted at $s$	&No  analysis  under noises\\
		\midrule
		\cite{yan2023distributed} &Local-PE &Fixed and strongly connected digraph & Mean-square convergence to zero\\
		\midrule
		This work &Cooperative PE &Time-varying and jointly connected digraph & Mean-square convergence to zero\\
		\bottomrule[1.2pt]
	\end{tabular}
	\caption{Comparison with existing distributed parameter estimation algorithms.}
	\label{tab:compare}
\end{table}

The remainder of this paper is organized as follows. First, Section~\ref{sec:form} presents some preliminaries on graph theory and formulates the problem of distributed parameter estimation, which is subject to the stochastic observation noises. The distributed estimator is proposed in Section~\ref{sec:algo}, with the convergence analysis carried out in Section~\ref{sec:analysis}.  Finally, we verify the established results through numerical examples in Section~\ref{sec:simulation} and conclude the paper in Section~\ref{sec:conclude}.

A preliminary version of this paper has been reported as \cite{yan2023distributed}. As we will discuss more later, in the current paper, we develop a different algorithm which is more advantageous for relaxing both the excitation condition and the size of messages exchanged among the sensors.  Moreover, this paper provides full proofs of all lemmas and theorems.

\textit{Notations}: For a vector $v,$ we denote by $v^i$ the $i$-th entry of it. Moreover, we denote by $v\geq 0$ (resp. $v<0$) if each entry of $v$ is nonnegative (resp. negative). On the other hand, for a matrix $M$, $[M]_i$ denotes the $i$-th row of it, while $[M]_i^j$ denotes its entry at the $i$-th row and $j$-th column. For a set of vectors $v_{i} \in \mathbb{R}^{m_{i}},$ the vector $\left[v_{1}^{T}, \ldots, v_{N}^{T}\right]^{T}$ is denoted by $\col(v_{1}, \ldots, v_{N}).$ A matrix is called doubly stochastic, if all of its entries are non-negative and each of its rows and columns sums to $1$.

\section{Problem Formulation}\label{sec:form}
In this section, we will first present some preliminaries on graph theory. After that, formulation of the distributed parameter estimation problem will be introduced.

\subsection{Preliminaries on graph theory}\label{sec:pre}
In this paper, we consider a time-varying weighted digraph $\mathcal{G}(k)=(\mathcal{V},\mathcal{E}(k), A(k))$. Here, $\mathcal{V} = {1, \ldots, n}$ represents the set of sensors, while $\mathcal{E}(k) \subseteq \mathcal{V} \times \mathcal{V}$ defines the set of edges. Additionally, $A(k) = [a_{ij}(k)]$ is the weighted adjacency matrix of $\mathcal{G}(k)$. The entries in $A(k)$ are constrained to non-negative values, and specifically, $a_{ij}(k) > 0$ if and only if sensor $i$ directly receives information from sensor $j$, i.e., when $(j, i) \in \mathcal{E}(k).$ Accordingly, the sets of in-neighbors and out-neighbors of sensor $i$ are respectively defined as 
\begin{equation}
	\begin{split}
		\mathcal{N}_i^+(k)&\triangleq\{j\in \mathcal{V}|(j,i)\in \mathcal{E}(k)\},\\\mathcal{N}_i^-(k)&\triangleq\{j\in \mathcal{V}|(i,j)\in \mathcal{E}(k)\}.
	\end{split}
\end{equation}
The (weighted) in-degree and (weighted) out-degree of sensor $i$ are further denoted as
\begin{equation}
	\begin{split}
		\mathrm{deg}_{i, in}(k)&=\sum_{j=1}^{n} a_{ij}(k),\;
		\mathrm{deg}_{i, out}(k)=\sum_{j=1}^{n} a_{ji}(k).
	\end{split}
\end{equation}
The digraph $\mathcal{G}(k)$ is called a \textit{weight-balanced digraph} if $\mathrm{deg}_{i, in}(k)=\mathrm{deg}_{i, out}(k)$ holds for all $i\in\mathcal{V}$. Moreover, for any $k_1,k_2\in\mathbb{N}$ with $k_1\leq k_2$, we define the union of digraphs over time interval $[k_1,k_2]$ as
\begin{equation}\label{eqn:joint}
	\begin{split}
		\mathcal{G}(k_1,k_2) &\triangleq \cup_{k =k_1}^{k_2} \mathcal{G}(k)=(\mathcal{V},\cup_{k =k_1}^{k_2}\mathcal{E}(k), \Sigma_{k =k_1}^{k_2}A(k)).
	\end{split}
\end{equation}
A sequence of edges $\left(i_1, i_2\right),\left(i_2, i_3\right), \ldots,\left(i_{\ell-1}, i_\ell\right)$  is called a directed path from sensor $i_1$ to sensor $i_\ell$. The digraph $\mathcal{G}(k_1,k_2)$ is said to be \textit{jointly strongly connected}, if for any $i, j \in \mathcal{V}$, there exists a directed path from $i$ to $j$. Notice that from \eqref{eqn:joint}, there is no guarantee that the edges of a path in  $\mathcal{G}(k_1,k_2)$ are followed
in a temporal order.
Therefore, we further introduce the notion of sequential dynamic paths as follows:

\begin{definition}[Sequential dynamic path]
	Consider any $i,j\in\mathcal{V}$. Given $k_1\leq k_2$, a sequential dynamic path from sensor $i$ to sensor $j$ over time interval $\left[k_1, k_2\right]$ is a sequence of edges $\left(i_{k_1}, i_{k_1+1}\right),\left(i_{k_1+1}, i_{k_1+2}\right), \ldots,\left(i_{k_2-1}, i_{k_2}\right)$ such that $i_{k_1}=i$, $i_{k_2}=j$, and $\left(i_{k}, i_{k+1}\right) \in \mathcal{E}(k)$ for all $k_1\leq k \leq k_2$. Moreover, we define the length of this path by $k_2-k_1$. Sensor $i$ and sensor $j$ are respectively the tail and head of this path. 
\end{definition}

Clearly, a sequential dynamic path can be viewed as a method of consecutively hopping from the path's tail to its head, with making only one hop at a time. 

\subsection{Distributed parameter estimation problem}
In this paper, we consider the problem of distributed parameter estimation in a network of $n$ sensors. At each time $k$, every sensor $i\in \{1, \ldots, n\}$ outputs a noisy measurement $y_{i}(k)\in\mathbb{R}$ determined by a $d$-dimensional regressor $\phi_i(k)\in\mathbb{R}^d$. They are related via the following stochastic linear regression model:
\begin{equation}\label{eqn:LRE}
	y_{i}(k)=\theta^\prime \phi_i(k)+w_{i}(k), \quad k \geq 0.
\end{equation}
Here, $\theta\in \mathbb{R}^{d}$ is the parameter to be estimated, and $w_{i}(k)\in\mathbb{R}$ is the independent and identically distributed (i.i.d.) Gaussian random noise with zero mean and covariance $R_i\geq0$. Notice that $R_i$ need not be known by any sensor.

\begin{remark}
	\rm	As reported in \cite{goodwin2014adaptive}, the input-output relationship of a large class of stochastic
	linear and nonlinear dynamical systems can be cast as \eqref{eqn:LRE}. Specifically, $\phi_i(k)$ denotes a vector that is either a linear or nonlinear function of 
	$\mathcal{U}_i(k)\triangleq\{u_i(0),u_i(1),\ldots,u_i(k-1)\}$ and $\mathcal{Y}_i(k)\triangleq\{y_i(0),y_i(1),\ldots,y_i(k-1)\}$, where $u_i(t)$ and $y_i(t)$ are respectively the input and output signals of the $i$-th subsystem at time $t$.
	We emphasize that the problem of estimating $\theta$ through \eqref{eqn:LRE} is fundamental in many applications such as system identification and adaptive control (see, e.g., \cite{aastrom2013adaptive}).
\end{remark} 

\begin{remark}
This paper considers the case where $\phi_i(k)$ is the excitation signal intentionally generated by the operator and contains no noises. On the other hand, we can also consider the case where $\phi_i(k)$ is another measured signal and can be contaminated by noises. In this case, we denote by $\tilde\phi_i(k)$ the actual excitation signal such that
\begin{equation*}
\begin{aligned}
\phi_i(k) &= \tilde\phi_i(k) + v_i(k),\\
y_i(k) &= \theta'\tilde\phi_i(k)+w_i(k),
\end{aligned}
\end{equation*}
where both $v_i(k)\in \mathbb{R}^d$ and $w_i(k)\in\mathbb{R}$ are i.i.d. Gaussian random noises with zero mean and bounded covariance. We thus have
\begin{equation*}
y_i(k) = \theta'(\phi_i(k)-v_i(k))+w_i(k) = \theta'\phi_i(k)+\tilde w_i(k),
\end{equation*}
where $\tilde w_i(k)\triangleq w_i(k)-\theta'v_i(k)$. It is clear that $\tilde w_i(k)$ is also an i.i.d. Gaussian random noise with zero mean and bounded covariance. Therefore, this dynamics exhibits the same form as \eqref{eqn:LRE}. As such, all the results in the paper remain valid in this case.
\end{remark}

The sensors aim to estimate $\theta$ from a stream of (noisy) measurable signals by exchanging information with each other. In this paper, we assume that the sensors communicate over a time-varying digraph $\mathcal{G}(k)=(\mathcal{V},\mathcal{E}(k),A(k))$.

\section{Estimation Algorithm Design}\label{sec:algo}
This section is devoted to proposing a distributed estimation algorithm. We shall employ the stochastic \textit{dynamic regressor extension and mixing} (DREM) algorithm, which transforms the problem of estimating a $d$-dimensional vector parameter to that of $d$ scalar ones.

\subsection{Stochastic dynamic regressor extension and mixing (DREM)}
The stochastic DREM algorithm is expressed by the following variables for each sensor $i\in\mathcal{V}$:
\begin{equation}\label{eqn:definition}  
\begin{split}
\Phi_{i}(k)&\triangleq\begin{bmatrix}
\left(\phi_{i}(k)\right)^\prime \\
\left(\phi_{i}(k-1)\right)^\prime \\
\vdots \\
\left(\phi_{i}(k-d+1)\right)^\prime
\end{bmatrix}\in\mathbb{R}^{d\times d}, \\\overline{y}_{i}(k) &\triangleq\adj(\Phi_{i}(k))\left[\begin{array}{c}
y_{i}(k) \\
y_{i}(k-1) \\
\vdots \\
y_{i}(k-d+1)
\end{array}\right]\in\mathbb{R}^{d},\\
\overline{w}_i(k)&\triangleq \adj(\Phi_{i}(k)) \begin{bmatrix}
w_{i}(k) \\
w_{i}(k-1)\\
\vdots \\
w_{i}(k-d+1)
\end{bmatrix}\in\mathbb{R}^{d},\\
\delta_{i}(k)&\triangleq \det(\Phi_{i}(k)),
\end{split}                      
\end{equation}
where we respectively denote by $\adj(\Phi_{i}(k))$ and $\det(\Phi_{i}(k))$ the adjugate matrix and the determinant of $\Phi_{i}(k)$. We further denote by $\overline{y}^\ell_{i}(k)$ and $\overline{w}^\ell_{i}(k)$ the $\ell$-th entries of $\overline{y}_{i}(k)$ and $\overline{w}_{i}(k)$, respectively. 

%The DREM algorithm was first proposed in \cite{aranovskiy2017performance} and recently reviewed in \cite{ortega2020new}. With the above variables, it decouples the $d$-dimensional estimation problem into $d$ scalar ones and allows each entry of the unknown parameter to be independently inferred. As reported in \cite{ortega2020new}, the DREM demonstrates decent performance in relaxing the excitation condition and guaranteeing asymptotic convergence in practical applications. However, notice that the traditional DREM algorithm works in deterministic systems (see, for example, \cite{aranovskiy2017performance,ortega2020new,pyrkin2019adaptive,yi2022conditions,matveev2021diffusion,bobtsov2022generation}). 
%
%Therefore, in order to accommodate the Guassian noises here, we shall first introduce subtle modifications to it. To this end, let us define
%\begin{equation}\label{eqn:noise}
%	\overline{w}_i(k)\triangleq \adj(\Phi_{i}(k)) \begin{bmatrix}
%		w_{i}(k) \\
%		w_{i}(k-1)\\
%		\vdots \\
%		w_{i}(k-d+1)
%	\end{bmatrix}.
%\end{equation}

Then, let us introduce the following lemma, which extends a result in \cite{aranovskiy2017performance} to stochastic scenarios:

\begin{lemma}\label{lmm:Y}
	Consider the network of sensors satisfying the stochastic linear regression model \eqref{eqn:LRE}. For each $\ell\in\{1,\ldots,d\},$ it holds for any $i\in\mathcal{V}$ that
	\begin{equation}\label{eqn:scalarLRE}
	\overline{y}_i^\ell(k) = \delta_{i}(k)\theta^\ell+\overline{w}_i^\ell(k),
	\end{equation}
	where $\theta^\ell$ is the $\ell$-th entry of the true parameter $\theta$.
\end{lemma}
\begin{pf*}{Proof.}
	%	We can show this immediately by writing \eqref{eqn:LRE} in a vector form as
	%	\begin{equation}\label{eqn:Y}
	%		\begin{split}
	%			\begin{bmatrix}
	%				y_{i}(k) \\
	%				y_{i}(k-1) \\
	%				\vdots \\
	%				y_{i}(k-d+1)
	%			\end{bmatrix}\!\!
	%			&=\!\!\begin{bmatrix}
	%				\left(\phi_{i}(k)\right)^\prime \\
	%				\left(\phi_{i}(k-1)\right)^\prime \\
	%				\vdots \\
	%				\left(\phi_{i}(k-d+1)\right)^\prime
	%			\end{bmatrix}\!\!\theta\! +\!\!\begin{bmatrix}
	%				w_{i}(k) \\
	%				w_{i}(k-1)\\
	%				\vdots \\
	%				w_{i}(k-d+1)
	%			\end{bmatrix}\!\!.
	%		\end{split}
	%	\end{equation}
	%	Then multiplying \eqref{eqn:Y} from the left by $\adj(\Phi_{i}(t))$, we have
	%	\begin{equation}
	%		\overline{y}_i(k) = \delta_{i}(k)\theta+\overline{w}_i(k).
	%	\end{equation}
	%	This holds because for any matrix $M\in\mathbb{R}^{d\times d}$, it follows that
	%	$
	%	\adj(M) M=\det(M)  I_{d}.
	%	$
	%	We thus arrive at the expression in \eqref{eqn:scalarLRE}.
	The proof follows similar arguments as in \cite{aranovskiy2017performance}. \hfill$\square$
\end{pf*}

We must point out that, due to inherent nature of the stochastic DREM, any (noisy) measurement is used for multiple times and the noise $\{\overline{w}_i(k)\}$ after transformation is no longer i.i.d. In the next section, we will propose a distributed estimation algorithm under such correlated noises. 

\subsection{The proposed algorithm}

So far, by leveraging the stochastic DREM, we have generated $d$ scalar ones as presented in \eqref{eqn:scalarLRE}. By combining it with the classical least-mean square (LMS) scheme, we are ready to propose our distributed estimation algorithm. Specifically, at any time $k > 0$, each sensor $i\in\mathcal{V}$ makes an estimation as 
outlined in Algorithm~\ref{alg:CTA}.

\begin{algorithm}[h!] 
	%1:\: Receive $\hat{\theta}_{j}(k)$ from all in-neighboring sensors $j\in\mathcal{N}^+_i(k)$.
	
	\textbf{for} $\ell\in\{1,2,...,d\}$ \textbf{do}
	\qquad \begin{enumerate}
		\item[1):] Combine the local estimates in neighborhood as
		
		\textit{(Combination)  } 
		\begin{equation}\label{eqn:resilientcomb}
			\begin{split}
				\bar{\theta}_i^\ell (k) =
				a_{ii}(k)\hat{\theta}_i^\ell (k)+\sum_{j\in\mathcal{N}^+_i(k)}a_{ij}(k)\hat{\theta}_j^\ell (k),
			\end{split}
		\end{equation}
		where $a_{ij}(k)$ is the $(i,j)$-th entry of the weighted adjacency matrix $A(k)$.
		\item[2):] Adapt the $\ell$-th entry of local estimate as
		
		\textit{(Adaptation)  }
		\begin{equation}\label{eqn:update}
			\begin{split}
				&\hat{\theta}^{\ell}_{i}(k+1)\\&\;=\bar{\theta}_i^\ell (k)
				+\frac{\alpha(k)\delta_{i}(k)}{\mu_{i}+\left(\delta_{i}(k)\right)^{2}}\left(\overline{y}^{\ell}_{i}(k)-\delta_{i}(k) \bar{\theta}_i^\ell (k)\right),
			\end{split}
		\end{equation}
		where $\mu_i>0$ and $\alpha(k)$ is a monotonically non-increasing stepsize. Additionally, $\alpha(k)$ satisfies the following conditions:
	\begin{equation}\label{eqn:alpha} 
	0<\alpha(k)\leq 1, \;\sum_{k=0}^\infty \alpha(k) = \infty, \;\sum_{k=0}^\infty \alpha^2(k)<\infty.
	\end{equation}
	\end{enumerate}
	\textbf{end for}

	%4:\: Transmit $\hat{\theta}_{i}(k+1)=\col(\hat{\theta}^{\ell}_{i}(k+1))$ to out-neighbors $j\in\mathcal{N}^-_i(k)$. \\
	\caption{CTA diffusion-based estimation algorithm under Gaussian observation noises}
	\label{alg:CTA}
\end{algorithm}

As in existing diffusion-based estimation algorithms (for example, \cite{xie2018analysis,tu2012diffusion,nosrati2015adaptive,takahashi2010diffusion}), in Algorithm~\ref{alg:CTA}, the update of each sensor involves two steps: combination step \eqref{eqn:resilientcomb} and adaptation step \eqref{eqn:update}.  However, different from such previous works, in the adaptation step, we process the new measurement by incorporating the stochastic DREM to the LMS scheme. Moreover, a sequence of time-varying stepsizes $\{\alpha(k)\}$ is adopted in \eqref{eqn:update}. This is different from the works \cite{xie2020convergence,xie2018analysis,abdolee2014diffusion}, in which the stepsizes are constant and only convergence to a neighborhood of the true parameter is achieved. We will show in Section~\ref{sec:analysis} that, by properly designing $\{\alpha(k)\}$, the proposed algorithm guarantees convergence to the true parameter in mean square. 

In Algorithm~\ref{alg:CTA}, we adopt a CTA algorithm. Later in Section~\ref{sec:ATC}, we will show how to extend Algorithm~\ref{alg:CTA} and obtain an ATC-type estimation scheme.

%We highlight that the proposed algorithm can be implemented in a fully distributed manner with limited information exchange among the sensors. 
%Notice that in Algorithm~\ref{alg:CTA}, the sensors
%transmit their local estimates $\hat{\theta}_i(k)$
%over the edges present in the network at that time.
%Each sensor makes updates in their local estimates
%with only its own and in-neighbors' information.
%The messages transmitted at each instant by any sensor
%is of size $d$. Compared to other existing algorithms,
%this is the minimal size. 
%For example, in \cite{yan2023distributed}, each sensor transmits both its local estimate
%and regressor to its neighbors, resulting in
%message size of $2d$. Moreover, in \cite{xie2020convergence}, in 
%addition to the estimated state, the error covariance matrix
%must be transmitted. This leads to a higher message size as $d^2+d$.

%As compared to the existing algorithms in \cite{lopes2008diffusion,cattivelli2009diffusion,xie2018analysis,tu2012diffusion,matveev2021diffusion,nosrati2015adaptive,takahashi2010diffusion,aranovskiy2017performance}, Algorithm~\ref{alg:CTA} incorporates the stochastic DREM to the well-known LMS algorithm. 	
%As will be shown in Section~\ref{sec:analysis}, this feature turns out to have a major impact on analyzing the stability of the linear time-varying systems that describe the behavior of estimation errors. 

\section{Performance Analysis}\label{sec:analysis}
This section analyzes the performance of Algorithm~\ref{alg:CTA}. We will show that if certain conditions on the network topology and regressors are met, each sensor can infer the true parameter in mean square.

\subsection{Cooperative persistent excitation (PE) condition}\label{sec:PE}

As observed from \eqref{eqn:update}, one factor that affects
convergence properties of the proposed estimator is the
determinant of the extended regressor $\delta_{i}(k),\;\forall i\in\mathcal{V}$. Since in the network of multiple sensors, it is difficult to guarantee the PE condition at each sensor's side, we shall relax the PE condition and introduce the following \textit{cooperative PE condition} that will be used in the rest of this paper. Note that this condition is imposed on the scalar regressor $\{\delta_i(k)\}$ rather than the original $\{\phi_i(k)\}$.
\begin{definition}[Cooperative PE condition]\label{def:localPE}
	The group of regressors $\{\delta_{i}(k)\}_{i\in\mathcal{V}}$ is said to satisfy the cooperative PE condition, if there exist $\omega>0$ and a finite time $T\in\mathbb{N}_{+}$ such that the following relation holds:
	\begin{equation}\label{eqn:PE}
	\sum_{t=k}^{k+T-1}\Big[\sum_{i=1}^n(\delta_i(\revise{t}))^2\Big] \geq \omega, \;\forall k.
	\end{equation} 
\end{definition}

The cooperative PE condition was first proposed in \cite{chen2013distributed}, where a consensus problem of a group of deterministic systems in the continuous-time domain is considered. Here, we make subtle modifications to accommodate the discrete-time stochastic signals as well as the scalar regressors in DREM. 

The cooperative PE condition is much
weaker than the conventional PE condition, since it is possible that none of the individual regressors meets the PE condition, but they collectively
satisfy the cooperative PE.  
Moreover, note that in most distributed
adaptive filters, such as those in \cite{abdolee2014diffusion,piggott2015stability,gharehshiran2013distributed}, it is required that the regressors are generated independently and statistically stationary. In contrast, the proposed cooperative PE condition removes this assumption by naturally generalizing the PE condition from a single sensor to a sensor network. Hence, it can be easily satisfied even in stochastic systems with feedback. Finally, in contrast to the cooperative stochastic information condition in \cite{xie2020convergence} and \cite{xie2018analysis}, which is established by taking the conditional expectation on all historical information, our condition is verified easily.

Under the cooperative PE condition \eqref{eqn:PE}, we can further conclude that, over any interval of length $T$, there exists at least one sensor in the network that is sufficiently excited. That is, for any $k\in\mathbb{N}$, there exist $t\in [k, k+T-1]$ and $j\in\{1,\ldots,n\}$ such that 
\begin{equation}\label{eqn:PE2}
(\delta_j(t))^2 \geq \underline{\omega}, 
\end{equation} 
where
$
\underline{\omega} \triangleq \omega/nT.
$

Notice that in \eqref{eqn:PE}, we propose the cooperative PE condition with respect to the scalar regressor $\delta_{i}(k)$. A question hence arises naturally: how is the excitation property of the scalar regressors related to that of the original regressor $\phi_i(k)$? To answer this question, we introduce the following proposition, the proof of which is given in Appendix~\ref{app:A}:
\begin{proposition}\label{prop:PE}
The cooperative PE condition \eqref{eqn:PE} holds, if and only if for any $k$, there exists a sensor $i\in\mathcal{V}$, $t\in[k,k+T-d]$, and $\omega_2>0$ such that
\begin{equation}\label{eqn:PE_2}
\sum_{\tau=t}^{t+d-1}\phi_i(\tau)\phi'_i(\tau)\geq \omega_2 I. 
\end{equation} 
\end{proposition}

\subsection{Assumptions}\label{sec:assumption}
As one might imagine, the estimation performance depends on both
information exchange in the communication network and informativeness of the signals measured
by various sensors. Therefore, we introduce some assumptions on them:

\begin{assumption}\label{assup:assumptions}
	\begin{enumerate}
		\item At any time $k$,  the sensors communicate over a weight-balanced digraph $\mathcal{G}(k)=(\mathcal{V},\mathcal{E}(k), A(k))$ such that $A(k)$ is doubly stochastic. Moreover, there exists $\underline{a}>0$ such that $a_{ij}(k)\geq \underline{a},$  for all $i\in\mathcal{V}$ and $ j\in \mathcal{N}^+_i(k)\cup\{i\}$. 
		\item There exists a finite time $h\in\mathbb{N}$ such that $\mathcal{G}(k,k+h) $ is jointly strongly connected for any $k\in\mathbb{N}$.
		\item The regressors of sensors are bounded and satisfy the cooperative PE condition \eqref{eqn:PE}.
		\item The stepsize $\{\alpha(k)\}$ is monotonically non-increasing and satisfies \eqref{eqn:alpha}.
	\end{enumerate}
\end{assumption}

Assumption~\ref{assup:assumptions} is commonly adopted in the literature and easily met in practice. To be specific, the first two assumptions pertain to the communication network. Informally speaking, Assumption 1.1 says that, at each time step, every sensor takes a convex combination of its own estimate and those of its in-neighbors by \eqref{eqn:resilientcomb}. Moreover, it assigns a substantial weight to each received information. Together with the doubly stochastic assumption on $A(k)$, it ensures that the information
from all sensors is mixed with equal weight in the long run. Assumption 1.2 further guarantees that every pair of sensors can
exchange information for infinite times throughout the execution. Notice that similar assumptions are also made in \cite{nedic2015distributed,lorenz2010conditions,yan2021resilient}.
The third assumption, on the other hand, refers to richness of the regressors that is necessary for consistently estimating the unknown parameter.
Finally, Assumption 1.4 is a standard rule on the stepsize and is widely used in stochastic approximation (\cite{li2010consensus,bianchi2013performance,pu2021sharp}). As will be seen, it plays an important role in guaranteeing that the estimation error
converges to zero under the stochastic noises.

\subsection{Derivation of the error equations}

Let us define the estimation error for any sensor $i\in\{1,\ldots,n\}$ and dimension $\ell \in \{1,\ldots,d\}$ as
\begin{equation}\label{eqn:error1}
\widetilde{\theta}^{\ell}_{i}(k) \triangleq \hat{\theta}^{\ell}_{i}(k)-\theta^\ell.
\end{equation}
In order to analyze the performance of Algorithm~\ref{alg:CTA}, we first study the dynamics of $\widetilde{\theta}^{\ell}_{i}(k)$.
Let us denote
\begin{equation}\label{eqn:error2}
	\begin{split}
		\widehat{\Theta}^{\ell}(k)&\triangleq\col(\hat{\theta}^{\ell}_{1}(k), \ldots, \hat{\theta}^{\ell}_{n}(k)),\\
		\overline{\Theta}^{\ell}(k)&\triangleq\col\left(\bar{\theta}_1^\ell (k) , \ldots, \bar{\theta}_n^\ell (k) \right), \\
		\widetilde{\Theta}^{\ell}(k)&\triangleq\col(\widetilde{\theta}^{\ell}_{1}(k), \ldots, \widetilde{\theta}^{\ell}_{n}(k)).
	\end{split}
\end{equation}
Moreover, let the vectors of outputs and noises be
\begin{equation}
	\begin{split}
		Y^{\ell}(k)&\triangleq\col(\overline{y}^\ell_{1}(k), \ldots, \overline{y}^\ell_{n}(k)) \in \mathbb{R}^n,\\
		W^{\ell}(k)&\triangleq\col(\overline{w}^\ell_{1}(k), \ldots, \overline{w}^\ell_{n}(k))\in \mathbb{R}^n,
	\end{split}
\end{equation}
and the matrix containing the scalar regressors be
\begin{equation}
	\Delta(k)\triangleq\operatorname{diag}\left(\delta_1(k), \ldots, \delta_n(k)\right)\in \mathbb{R}^{n\times n}.
\end{equation}
We then obtain the vector representation of \eqref{eqn:scalarLRE}:
\begin{equation}\label{eqn:yMatrix}
	Y^{\ell}(k)=\Delta(k) \Theta^{\ell}+ W^{\ell}(k),
\end{equation}
where $$\Theta^{\ell}\triangleq\1_n\theta^{\ell} \in \mathbb{R}^n.$$ Furthermore, the two steps \eqref{eqn:resilientcomb} and \eqref{eqn:update} in Algorithm~\ref{alg:CTA} can be expressed as 
\begin{equation*}
	\begin{split}
		\overline{\Theta}^{\ell}(k) &=A(k) \widehat{\Theta}^{\ell}(k), \\
		\widehat{\Theta}^{\ell}(k+1) &=\overline{\Theta}^{\ell}(k) +\alpha(k)L(k)\left(Y^{\ell}(k)-\Delta(k) \overline{\Theta}^{\ell}(k)\right),
	\end{split}
\end{equation*}
where $A(k)$ is the weighted adjacency matrix and \begin{equation}
	\begin{split}\label{eqn:L}
		L(k)&\triangleq\operatorname{diag}\left(\frac{\delta_1(k)}{\mu_1+\left(\delta_1(k)\right)^2}, \ldots, \frac{\delta_n(k)}{\mu_n+\left(\delta_n(k)\right)^2}\right).
	\end{split}
\end{equation}
Combining the above equations with \eqref{eqn:yMatrix}, we can write the update in the estimation error vector as
\begin{equation}\label{eqn:tildeTheta}
	\begin{aligned}
		\widetilde{\Theta}^{\ell}(k+1) =\;&A(k) \widehat{\Theta}^{\ell}(k)-\Theta^{\ell} +\alpha(k)L(k)W^\ell(k)\\
		&+\alpha(k)L(k)\left(\Delta(k) \Theta^{\ell}-\Delta(k) A(k) \widehat{\Theta}^{\ell}(k)\right).
	\end{aligned}
\end{equation}
%Now, let us consider any entry of $\widetilde{\Theta}^{\ell}(k+1)$, namely, $\widetilde{\theta}^{\ell}_{i}(k+1),\forall i\in\{1,\ldots,n\}$. Combining \eqref{eqn:tildeTheta} with \eqref{eqn:update} and \eqref{eqn:delta}, it follows that
%\begin{equation}\label{eqn:single}
%\begin{split}
%\widetilde{\theta}^{\ell}_{i}(k+1)=
%&\Big(1-\frac{\alpha(k)(\delta_i(k))^2}{\mu_i+(\delta_i(k))^2}\Big)\sum_{j\in\mathcal{N}^+_i(k)\cup\{i\}}a_{ij}(k)\widetilde{\theta}_j^\ell (k)\\&+\frac{\alpha(k)\delta_i(k)}{\mu_i+(\delta_i(k))^2}\overline{w}^\ell_{i}(k).
%\end{split}
%\end{equation}
%Notice that from this equation, one can verify that when $\delta_i(k)=0$, it follows that
%\begin{equation}
%\widetilde{\theta}^{\ell}_{i}(k+1) = \sum_{j\in\mathcal{N}^+_i(k)\cup\{i\}}a_{ij}(k)\widetilde{\theta}_j^\ell (k).
%\end{equation}
%This agrees with \eqref{eqn:update} where sensor $i$ adapts the local estimate only when $\delta_i(k)\neq 0$. 
Since $A(k) \Theta^{\ell}=\Theta^{\ell}$, we conclude that the dynamics of estimation error as
\begin{equation}\label{eqn:error}
	\begin{aligned}
		\widetilde{\Theta}^{\ell}(k+1)  &=\left(I-\alpha(k)L(k) \Delta(k)\right) A(k) \widetilde{\Theta}^{\ell}(k)  \\&\quad+\alpha(k)L(k)W^\ell(k)\\&=G(k) A(k) \widetilde{\Theta}^{\ell}(k) +\alpha(k)L(k)W^\ell(k),
	\end{aligned}
\end{equation}
where
\begin{equation}\label{eqn:G}
	\begin{split}
		G(k) &\triangleq I-\alpha(k)L(k) \Delta(k).
	\end{split}
\end{equation}
Note that in this iterative equation, 
the matrices $A(k)$ and
$G(k)$ respectively represent the communication topology and the information content of the regressors.

\subsection{Convergence analysis}
So far, we have obtained the dynamics of estimation error \eqref{eqn:error}. In what follows, we shall prove that this error converges to zero in mean square for any $\ell \in \{1,\ldots,d\}$. In the rest of this section, we omit the superscript $\ell$ in the 
notation appearing in $\Theta^{\ell}$ and so on. Note that the sizes of these vectors are the same for all $\ell$.

Let us consider the following Lyapunov function candidate:
\begin{align}
	V(k) &\triangleq \widetilde{\Theta}^\prime(k) \widetilde{\Theta}(k) \notag \\&=\widetilde{\Theta}^\prime(k)J \widetilde{\Theta}(k)+\widetilde{\Theta}^\prime(k)(I-J) \widetilde{\Theta}(k) \notag\\&=V_1(k)+V_2(k),\label{eqn:Lyapunov}
\end{align}
where 
\begin{equation}\label{eqn:Jdef}
	J \triangleq \frac{\1_n\1_n^\prime}{n},
\end{equation}
and
\begin{equation}\label{eqn:V}
	\begin{split}
		V_1(k)&\triangleq\widetilde{\Theta}^\prime(k)J \widetilde{\Theta}(k),\\
		V_2(k)&\triangleq \widetilde{\Theta}^\prime(k)(I-J) \widetilde{\Theta}(k).
	\end{split}
\end{equation}
It is easy to check that 
\begin{equation}\label{eqn:J}
	J^\prime J =J,\;(I-J)^\prime (I-J) =I-J.
\end{equation}

We next introduce the following lemmas related to the two matrices $A(k)$ and $G(k)$ in \eqref{eqn:error}.

\begin{lemma}\label{lmm:variance}
	For $x\in\mathbb{R}^d$, suppose that $x' (I-J) x\leq  c x'  x$ holds with
		\begin{equation}\label{eqn:c}
		0<c<\frac{1}{n+1},
		\end{equation}
	\revise{and $J$ defined in \eqref{eqn:Jdef}.} Then, it holds that either $x\geq 0$ or $x<0$.
	Moreover, for any $i\in\{1,\ldots,n\}$, it follows that $$(x^i)^2\geq \frac{1}{n} \Big(1-\sqrt{\frac{cn}{1-c}}\Big)^2x' J x,$$ where $x^i$ is the $i$-th entry of $x$.
\end{lemma}
\begin{pf*}{Proof.}
	By definition of the matrix $J$, it is easy to verify by \eqref{eqn:J} that
	\begin{equation}
	\begin{split}
	x'Jx  &= n\nu^2,\;x' (I-J) x = \sum_{i=1}^n (x^i-\nu)^2,
	\end{split}
	\end{equation}
	where
	$
	\nu \triangleq \frac{1}{n} \sum_{i=1}^n x^i.
	$ Since $x'  x = x' (I-J) x + x' J x $ and $0<c<\frac{1}{n+1}<1$, we conclude that
	\begin{equation}
	x'(I-J)x \leq \frac{c}{1-c}x'Jx.
	\end{equation}
	Namely,
	\begin{equation}
	\sum_{i=1}^n (x^i-\nu)^2 \leq \frac{cn}{1-c} \nu^2.
	\end{equation}
	Therefore, for any $i\in\{1,\ldots,n\}$, it follows that 
	\begin{equation}
	(x^i-\nu)^2 \leq \frac{cn}{1-c} \nu^2.
	\end{equation}
	By \eqref{eqn:c}, it holds $\frac{cn}{1-c}<1$. Hence, one of the following statements holds:
	\begin{enumerate}
		\item $\nu<0$ and $\nu+\sqrt{\frac{cn}{1-c}}\nu\leq x^i \leq \nu-\sqrt{\frac{cn}{1-c}}\nu<0$;
		\item $\nu>0$ and $0\leq \nu-\sqrt{\frac{cn}{1-c}}\nu\leq x^i \leq \nu+\sqrt{\frac{cn}{1-c}}\nu$.
	\end{enumerate}
	In either case, it follows that
	\begin{equation*}
	(x^i)^2 \geq \Big(1-\sqrt{\frac{cn}{1-c}}\Big)^2\nu^2 = \frac{1}{n} \Big(1-\sqrt{\frac{cn}{1-c}}\Big)^2x' J x.
	\end{equation*}
	We therefore complete the proof.\hfill$\square$
\end{pf*}

\begin{lemma}\label{lmm:Ax}
	Given any $x\in\mathbb{R}^n$, for any doubly stochastic matrix $A$, it follows that
	\begin{equation}
		\begin{split}
			x' x&\geq  x' A'A x,\\
			x' (I-J)  x&\geq  x' A'(I-J) A x.
		\end{split}
	\end{equation}
\end{lemma}
\begin{pf*}{Proof.}
	Notice that the matrix $A'A$ is symmetric and any of its eigenvalue is between $0$ and $1$ (\hspace{1pt}\cite{horn2012matrix}). We thus have
	\begin{equation}
		x' x-  x' A'A x = x'(I-A'A)x\geq 0.
	\end{equation}
	On the other hand, it follows that
	$
		A'J' = J',\; JA = J.
$
	Therefore, we have 
	\begin{equation}
		A'JA = A'J'JA = J'J =J.
	\end{equation}
	One thus concludes that
	\begin{equation}
		\begin{split}
			&x'(I-J) x - x' A'(I-J) A x \\&= x'((I-J)-A'(I-J)A)x \\&= x'(I-A'A)x\geq 0.
		\end{split}
	\end{equation}
	We hence complete the proof.
	$\hfill\square$
\end{pf*}

\begin{lemma}\label{lmm:Gx}
	Given any $x\in\mathbb{R}^n$, it follows that
	\begin{equation}
		\begin{split}
			x' x\geq  x' G'(k)G(k) x,\;\forall k,
		\end{split}
	\end{equation}
	where $G(k)$ is defined in \eqref{eqn:G}. Moreover, if there exists at least one sensor, labeled as $i\in\mathcal{V}$, that is excited at time $k$ such that $(\delta_{i}(k))^2\geq \underline{\omega}$, then it holds that\footnote{In this paper, with slight abuse in the statement, we say that a sensor $i$ is excited at time $k$ if $(\delta_{i}(k))^2\geq \underline{\omega}$. However, we should note that $(\delta_{i}(k))^2$ contains the historical information in the interval $[k-d+1,k],$ as observed from \eqref{eqn:definition}.}
	\begin{equation}
		\begin{split}
			x' x-  x' G'(k)G(k) x \geq \zeta_i(k)(x^i)^2,
		\end{split}
	\end{equation}
	where
	\begin{equation}\label{eqn:zeta}
		\zeta_i(k) \triangleq \left(2-\frac{\alpha(k)\underline{\omega}}{\mu_i+\underline{\omega}}\right)\frac{\alpha(k)\underline{\omega}}{\mu_i+\underline{\omega}}\in (0,1).
	\end{equation}
\end{lemma}
\begin{pf*}{Proof.}
	By \eqref{eqn:L} and \eqref{eqn:G}, we can calculate that
	\begin{equation*}
		G(k) \!=\! \operatorname{diag}\left(1-\frac{\alpha(k)\left(\delta_1(k)\right)^2}{\mu_1+\left(\delta_1(k)\right)^2}, \ldots, 1-\frac{\alpha(k)\left(\delta_n(k)\right)^2}{\mu_n+\left(\delta_n(k)\right)^2}\right)\!.
	\end{equation*}
	The proof is thus straightforward by noting that
	\begin{equation}\label{eqn:difference}
		\begin{split}
			&x' x-  x' G'(k)G(k) x \\&= \sum_{i=1}^{n}(x^i)^2-\sum_{i=1}^n(x^i)^2\left(1-\frac{\alpha(k)\left(\delta_i(k)\right)^2}{\mu_i+\left(\delta_i(k)\right)^2}\right)^2.
		\end{split}
	\end{equation}
	For simplicity, we denote $$\xi_i(k)\triangleq \frac{\alpha(k)\left(\delta_i(k)\right)^2}{\mu_i+\left(\delta_i(k)\right)^2}\in[0,1).$$
	It thus follows from \eqref{eqn:difference} that
	\begin{equation*}
		\begin{split}
			&x' x-  x' G'(k)G(k) x =\sum_{i=1}^{n}\big((2-\xi_i(k))\xi_i(k)\big)(x^i)^2\geq 0.
		\end{split}
	\end{equation*}
	On the other hand, it is not difficult to verify that $(2-\xi_i(k))\xi_i(k)$ increases monotonically with $(\delta_{i}(k))^2$.
	Therefore, if there exists a sensor $i\in\mathcal{V}$ such that $(\delta_{i}(k))^2\geq \underline{\omega}$, we obtain from the above equation that
	\begin{equation*}
		x' x-  x' G'(k)G(k) x \geq \left(2-\frac{\alpha(k)\underline{\omega}}{\mu_i+\underline{\omega}}\right)\frac{\alpha(k)\underline{\omega}}{\mu_i+\underline{\omega}}(x^i)^2.
	\end{equation*}
	The proof is thus complete. \hfill$\square$
\end{pf*}

In view of Lemmas~\ref{lmm:Ax} and \ref{lmm:Gx}, both matrices $A(k)$ and $G(k)$ can contribute to decrease the Lyapunov function $V(k)$ in \eqref{eqn:Lyapunov}. Next, we shall investigate further on the communication among sensors.

To see this, for any $k,s \in \mathbb{N}$ with $k <s,$ let us denote the transition matrix obtained from $A(k)$ as
\begin{equation}
	\begin{split}
		\Phi_A(k, s)&\triangleq A(s) A(s-1) \cdots A(k+1) A(k).
	\end{split}
\end{equation}
With respect to it, we introduce the following result:

\begin{lemma}\label{lmm:topo}
	Suppose that Assumption \ref{assup:assumptions}(2) holds. There exists a finite time $H\leq nh$ such that for any $k\in\mathbb{N}$, the following statements hold:
	\begin{enumerate}
		\item For any $i,j\in\{1,\ldots,n\}$, there exists a sequential dynamic path from sensor $i$ to sensor $j$ over the time interval $[k,k+H]$;
		\item For any $\tau\geq H$, each entry of the transition matrix $\Phi_A(k, k+\tau)$ is lower bounded by $\underline{a}^H$. That is, 
		\begin{equation}
			[\Phi_A(k, k+\tau)]^i_j \geq \underline{a}^H, \;\forall i,j. 
		\end{equation}
	\end{enumerate}
\end{lemma}
\begin{pf*}{Proof.}
	Following similar proof of \cite[Lemma~1]{nedic2009distributed}, we can establish the first statement. Moreover, it holds for any $i$ and $j$ that \begin{equation}
		[\Phi_A(k, k+H)]_j^i  \geq \underline{a}^H.
	\end{equation}
	That is, each entry of $\Phi_A(k, k+H)$ is lowered bounded by $\underline{a}^H$. 
	Because $A(t)$ is always stochastic, for each $\tau\geq H$, any entry of $\Phi_A(k, k+\tau+1)$ is a convex combination of certain entries of  $\Phi_A(k, k+\tau)$. Therefore, it is also lower bounded by $\underline{a}^H$.
\hfill$\square$
\end{pf*}

For notational simplicity, we define
\begin{equation}\label{eqn:B}
	\bar{T}\triangleq  \max(T,H),
\end{equation} 
where $T$ and $H$ are respectively given in \eqref{eqn:PE} and Lemma~\ref{lmm:topo}.
Therefore,  given any $k\in\mathbb{N}$, there exists a sequential dynamic path between any pair of sensors over the interval $[k,k+\bar{T}]$. Moreover, at least one sensor is excited at some instant within this interval.

The last lemma given below will also be used in proving the main result. 
\begin{lemma}[\hspace{1pt}{\cite{polyak1987introduction}}]\label{lmm:converge}
	Let $\{u(k)\},\{p(k)\}$ and $\{q(k)\}$ be real sequences such that 
	\begin{equation}\label{eqn:lemma6}
		u(k+1) \leq(1-q(k)) u(k)+p(k).
	\end{equation}
	Suppose that the following conditions hold:
	\begin{enumerate}
		\item For any $k\in\mathbb{N}$, it holds that $p(k) \geq 0$;
		\item The stepsize satisfies that $0< q(k) \leq 1$ and $\sum_{k=0}^{\infty} q(k)=\infty$;
		\item $\lim\limits_{k \to \infty}\frac{p(k)}{q(k)} = 0$.
	\end{enumerate}
	If $u(k)\geq 0$ for $k\in\mathbb{N}$, then $u(k)$ asymptotically converges to zero, namely, $\lim _{k \rightarrow \infty} u(k) =0.$
\end{lemma}

With these preparations above, we are now ready to provide the main theorem. 
\begin{theorem}\label{thm:converge}
	Consider the network of sensors satisfying the stochastic linear regression model \eqref{eqn:LRE}.
	Suppose that Assumption~\ref{assup:assumptions} holds. By performing Algorithm~\ref{alg:CTA}, it follows that
	\begin{align}
		\lim\limits_{k\to \infty} &\mathbb{E}[\widetilde{\Theta}'(k)\widetilde{\Theta}(k)]=0, \label{eqn:L2}
	\end{align}
	where $\widetilde{\Theta}(k)$ is given by \eqref{eqn:error1} and \eqref{eqn:error2}. That is, each sensor infers the true parameter $\theta$ in the mean square sense. 
\end{theorem}

%\begin{remark}\label{rmk:proof}
%	\rm Before formally proving Theorem~\ref{thm:converge}, we introduce our ideas for making it.
%	Intuitively, the estimation performance is guaranteed by both the adaptation and combination steps in Algorithm~\ref{alg:CTA}. Specifically, the adaptation step ensures that the excited sensors move towards the true parameter. On the other hand, in the combination step, each sensor performs a consensus algorithm to fairly mix the local information in its neighborhood, in turn reducing the variance among all sensors. As will be seen later, these respectively correspond to Scenarios I and II in our proof, where certain properties of matrices $G(k)$ and $A(k)$ play a key role.
%\end{remark}

\begin{pf*}{Proof.}
Here is a brief outline of the proof. In \eqref{eqn:Lyapunov}, we decompose the Lyapunov function $V(k)$ into two components, namely $V_1(k)$ and $V_2(k)$. In what follows, we will analyze two distinct scenarios, distinguished by the relative significance of $V_1(k)$ and $V_2(k)$. In the first scenario, when $V_2(k)$ is smaller than a specified threshold, indicating that $V_1(k)$ dominates, the reduction of $V(k)$ primarily hinges on the adaptation step, where the matrix $G(k)$ plays a pivotal role. On the other hand, in the second scenario, when $V_2(k)$ takes precedence, we demonstrate that the matrix $A(k)$ within the combination step operates to diminish the value of $V(k)$.
	
Specifically, in this proof, we shall show that the estimation error converges in mean square. This will be done by employing Lemma~\ref{lmm:converge} and establishing that the three conditions there are satisfied for the Lyapunov function candidate $\mathbb{E}[V(k)]$ defined in \eqref{eqn:Lyapunov}. 
	To this end, let us recall the error dynamics \eqref{eqn:error}. For simplicity, we denote
	\begin{equation}\label{eqn:M}
		M(k) \triangleq G(k)A(k).
	\end{equation}
	For any $k,s \in \mathbb{N}$, let the transition matrix for $M(k)$ be
\begin{equation}\label{eqn:Pi}
		\begin{split}
			\Phi_M(k,s) &\triangleq M(s)M(s-1)\cdots M(k+1)M(k), \; k<s,\\
			\Phi_M(k,k) & \triangleq I.
		\end{split}
	\end{equation}
	Then, it follows from \eqref{eqn:error} that
	\begin{equation}\label{eqn:tTheta}
		\begin{split}
			&\widetilde{\Theta}(k+\bar{T}+1) \\&=\Phi_M(k,k+\bar{T}) \widetilde{\Theta}(k) +\!\sum_{t=k}^{k+\bar{T}}\alpha(t)\Phi_M(t,k+\bar{T})L(t)W(t),
		\end{split}
	\end{equation}
	where $\bar{T}$ is given in \eqref{eqn:B}. 
	
	%(1) \textit{Mean-square convergence}
	
	%In the first part, we show that \eqref{eqn:L2} holds. To this end, 
	
	By \eqref{eqn:definition}, for any $t\in [k,k+\bar{T}]$, $W(t)$ is correlated to $\{W(\tau)\}$ for $\tau\in[\max(k,t-d+1),\min(t+d-1,k+\bar{T})]$. Moreover, as assumed, $W(t)$ is zero-mean and bounded in covariance. Therefore, we conclude from \eqref{eqn:tTheta} that a constant $0\leq \Gamma<\infty$ exists such that
	\begin{equation}
		\begin{split}
			&\mathbb{E}[V(k+\bar{T}+1)]\\&\leq  \mathbb{E}[\widetilde{\Theta}'(k)\Phi_M'(k,k+\bar{T})\Phi_M(k,k+\bar{T})\widetilde{\Theta}(k)]\\&\quad+ \Gamma \sum_{t=k}^{k+\bar{T}}\sum_{\tau=\max(k,t-d+1)}^{\min(t+d-1,k+\bar{T})}\alpha(t)\alpha(\tau).
		\end{split}
	\end{equation}
	Since $\{\alpha(k)\}$ is monotonically non-increasing, we have
	\begin{equation}\label{eqn:errmean}
		\begin{split}
			&\mathbb{E}[V(k+\bar{T}+1)]\\&\leq  \mathbb{E}[\widetilde{\Theta}'(k)\Phi_M'(k,k+\bar{T})\Phi_M(k,k+\bar{T})\widetilde{\Theta}(k)]+ C\alpha^2(k),
		\end{split}
	\end{equation}
	where $C\triangleq \Gamma(2d+1)(\bar{T}+1)$.
	
	To analyze the convergence of \eqref{eqn:errmean}, we especially focus on the first term of the right-hand side of \eqref{eqn:errmean}.
	To this end, given any $0<r<\min(1,4/n)$, let us denote $$c\triangleq \frac{(1-\sqrt{nr})^2}{n+(1-\sqrt{nr})^2}\in(0,1).$$ 
We will find bounds on this term by examining two scenarios contingent on the size of $V_2(k)$ at each time step $k$. Recall that $V_1(k)$ and $V_2(k)$ are given in \eqref{eqn:V}.
	
	\textbf{Scenario I}: At time $k$, it holds $V_2(k)\leq c V(k).$ \\
	
	In this case, it holds that
	\begin{equation}\label{eqn:barV}
		V_1(k)\geq (1-c)V(k).
	\end{equation}
	Moreover, it is easy to check that
	\begin{equation}
		\begin{split}
\frac{cn}{1-c}=(1-\sqrt{nr})^2<1.
		\end{split}
	\end{equation}
	This indicates that $0<c<\frac{1}{n+1}$. In view of Lemma~\ref{lmm:variance}, either $\widetilde{\Theta}(k)\geq 0$ or $\widetilde{\Theta}(k)<0$ holds. Further, we have
	\begin{equation}\label{eqn:rV}
		\begin{split}
			(\widetilde{\theta}_i(k))^2 &\geq \frac{1}{n} \Big(1-\sqrt{\frac{cn}{1-c}}\Big)^2V_1(k) \\&= rV_1(k)\geq r(1-c)V(k),
		\end{split}
	\end{equation}
	where the last inequality holds by \eqref{eqn:barV}. That is, for each sensor, the square of its estimation error is lower bounded by $rV_1(k)$ in this scenario. 
	
	Then, let us define a sequence $\{x(t)\}_{t\in[k,k+\bar{T}]}$ as follows:
	\begin{equation}\label{eqn:x}
		\begin{split}
			x(k) &\triangleq \widetilde{\Theta}(k),\\
			x(t+1) &\triangleq M(t)x(t),\forall t\in[k,k+\bar{T}],
		\end{split}
	\end{equation}
	where $M(t)$ is defined in \eqref{eqn:M}. Note that by Assumption~\ref{assup:assumptions}.3, at least one sensor, labeled as $i$, will be excited at least once within the interval $[k,k+\bar{T}]$. Without loss of generality, suppose that it is excited at time $k_0\in[k,k+\bar{T}]$. 
	By \eqref{eqn:G}, at any $t\in[k,k_0-1],$ it holds
	$
		M(t)=A(t).
	$
	As such, we have
	$
	x(t+1) = A(t)x(t), \forall t\in[k,k_0-1].
	$
	By virtue of Lemma~\ref{lmm:Ax}, it is straightforward to see that
	\begin{equation}\label{eqn:k0}
		x'(k_0)x(k_0) \leq x'(k)x(k) = V(k).
	\end{equation}
	Moreover, since $A(t)$ is stochastic, for each $j\in\{1,\ldots,n\}$, $x^j(t+1)$ is a convex combination of all $x^i(t)$, where $i\in\mathcal{N}^+_j(t)\cup\{j\}$. Therefore, it follows that
	\begin{equation}
		\begin{split}
			x^i(k_0) &\geq \min_{j\in\{1,\ldots,n\}} x^j(k_0)\\&\geq \min_{j\in\{1,\ldots,n\}} x^j(k) =  \min_{j\in\{1,\ldots,n\}} \widetilde{\theta}_j(k).
		\end{split}
	\end{equation}
	Similarly, we have
	$
	x^i(k_0)\leq \max_{j\in\{1,\ldots,n\}} \widetilde{\theta}_j(k).
	$
	With similar arguments, one can also obtain that
	\begin{equation}
		\min_{j\in\{1,\ldots,n\}}\widetilde{\theta}_j(k) \leq A(k_0)x^i(k_0)\leq  \max_{j\in\{1,\ldots,n\}} \widetilde{\theta}_j(k).
	\end{equation}
	Since each entry of $\widetilde{\Theta}(k)$ has the same sign, one concludes from \eqref{eqn:rV} that
	\begin{equation*}
		(A(k_0)x^i(k_0))^2\geq \min_{j\in\{1,\ldots,n\}} (\widetilde{\theta}_j(k))^2 \geq r(1-c)V(k).
	\end{equation*}
	
	Next, as sensor $i$ is excited at time $k_0$, it can be obtained from Lemma~\ref{lmm:Gx} that
	\begin{equation*}
		\begin{split}
			&x'(k_0+1)x(k_0+1)= x'(k_0)A'(k_0)G'(k_0)G(k_0)A(k_0)x(k_0)\\&\leq x'(k_0)A'(k_0)A(k_0) x(k_0)-
			r(1-c)\zeta_i(k)V(k)\\&\leq x'(k_0) x(k_0)-
			r(1-c)\zeta_i(k)V(k)\\&\leq V(k)-
			r(1-c)\zeta_i(k)V(k), 
		\end{split}
	\end{equation*}
	where $\zeta_i(k)$ is defined in \eqref{eqn:zeta}, the second inequality holds by Lemma~\ref{lmm:Ax} and the last inequality holds by \eqref{eqn:k0}. 
	
	Recalling Lemmas~\ref{lmm:Ax} and \ref{lmm:Gx}, we can further verify 
	\begin{equation}\label{eqn:scenario1}
		\begin{split}
			&x'(k+\bar{T}+1)x(k+\bar{T}+1)\leq x'(k_0+1)x(k_0+1) \\&\leq V(k)-
			r(1-c)\zeta_i(k)V(k)\\&\leq V(k)-
			r(1-c)\underline\zeta(k)V(k),
		\end{split}
	\end{equation}
	where \begin{equation}\label{eqn:zeta2}
		\underline\zeta(k)\triangleq \min_{j\in\mathcal{V}}\zeta_i(k)=\left(2-\frac{\alpha(k)\underline{\omega}}{\bar\mu+\underline{\omega}}\right)\frac{\alpha(k)\underline{\omega}}{\bar\mu+\underline{\omega}}\in(0,1),
	\end{equation}
	and $\bar\mu\triangleq \max_j(\mu_j)$.
	
	In view of \eqref{eqn:x}, we know that
	\begin{equation}
		\begin{split}
			&x'(k+\bar{T}+1)x(k+\bar{T}+1) \\&= \widetilde{\Theta}'(k)\Phi_M'(k,k+\bar{T})\Phi_M(k,k+\bar{T})\widetilde{\Theta}(k).
		\end{split}
	\end{equation}
	By \eqref{eqn:errmean}, we finally conclude that
	\begin{equation}\label{eqn:case1}
		\begin{split}
			\mathbb{E}[V(k+\bar{T}+1)]\leq (1-\gamma_1(k))\mathbb{E}[V(k)]+ C\alpha^2(k),
		\end{split}
	\end{equation}
	where 
	\begin{equation}\label{eqn:gamma1}
		\gamma_1(k) \triangleq 
		r(1-c)\underline\zeta(k)\in(0,1).
	\end{equation}

	\textbf{Scenario II}: At time $k$, it holds $V_2(k)> c V(k)$.\\
	
	We prove that $V(k)$ also decreases in mean in this scenario. First note that by definition in \eqref{eqn:M},
	\begin{equation}
		M(t) = G(t)A(t)=(I-\alpha(t)L(t) \Delta(t))A(t), \;\forall t.
	\end{equation}
	By \eqref{eqn:Pi}, we rewrite $\Phi_M'(k,k+\bar{T})\Phi_M(k,k+\bar{T})$ as
	\begin{equation}\label{eqn:Theta}
		\begin{split}
			&\Phi_M'(k,k+\bar{T})\Phi_M(k,k+\bar{T})\\&= (A(k+\bar{T})\cdots A(k))'A(k+\bar{T})\cdots A(k)+Q_\alpha(k)\\&=\Phi_A'(k, k+\bar{T})\Phi_A(k, k+\bar{T})+Q_\alpha(k),
		\end{split}
	\end{equation}
	where $Q_\alpha(k)\triangleq\Phi_M'(k,k+\bar{T})\Phi_M(k,k+\bar{T})-\Phi_A'(k, k+\bar{T})\Phi_A(k, k+\bar{T})$ is the remainder. Since $\lim_{k \to \infty} \alpha(k) = 0,$ it is easy to see that $M(k)$ converges to $A(k)$ and thus $\Phi_M(k,k+\bar{T})$ converges to $\Phi_A(k, k+\bar{T})$ as $k\to\infty$. Therefore, we conclude that
	\begin{equation}\label{eqn:Q}
		\lim_{k \to \infty} Q_\alpha(k) = 0.
	\end{equation}
	On the other hand, by virtue of Lemma~\ref{lmm:topo}, it follows that
	\begin{equation}
		[\Phi_A(k, k+\bar{T})]_j^i \geq \underline{a}^H. 
	\end{equation}
	Let us denote $m\triangleq n \underline{a}^H$.
	Then, there exists $P(k)\in\mathbb{R}^{n\times n}$ such that
	\begin{equation}
		\Phi_A(k, k+\bar{T}) = m J + P(k),
	\end{equation}
	where
	\begin{equation}
		\begin{split}
			P(k)\1_n = (1-m)\1_n, \; \1_n' P(k)= (1-m)\1_n'.
		\end{split}
	\end{equation}
	Hence, let us define a matrix $\tilde{P}(k)$ as
	\begin{equation}
		\tilde{P}(k) \triangleq \frac{1}{1-m}P(k).
	\end{equation}
	Obviously, $\tilde{P}(k)$ is doubly stochastic with $\tilde{P}(k)\1_n = \1_n, \; \1_n' \tilde{P}(k)= \1_n'.$ Hence, for any $x\in\mathbb{R}^n$, we can obtain that
	\begin{equation}\label{eqn:P}
		\begin{split}
			&x'\Phi_A'(k, k+\bar{T})(I-J)^2\Phi_A(k, k+\bar{T}) x \\&= x' (m J + P(k))'(I-J)^2(m J + P(k))x\\&= x' P(k)'(I-J)P(k)x\\&=(1-m)^2 x' \tilde{P}(k)'(I-J)\tilde{P}(k)x.
		\end{split}
	\end{equation}
	
	\vspace{-15pt}
	
	As such, it can be obtained from \eqref{eqn:Theta} that
	\begin{equation*}
		\begin{split}
			&\widetilde{\Theta}'(k)\Phi_M'(k,k+\bar{T})\Phi_M(k,k+\bar{T})\widetilde{\Theta}(k) \\&=\widetilde{\Theta}'(k)\Phi_A'(k, k+\bar{T})\Phi_A(k, k+\bar{T})\widetilde{\Theta}(k)\\&\hspace{1cm}+\widetilde{\Theta}'(k)Q_\alpha(k)\widetilde{\Theta}(k)\\&=\widetilde{\Theta}'(k)J\widetilde{\Theta}(k)+\widetilde{\Theta}'(k)Q_\alpha(k)\widetilde{\Theta}(k)\\&\hspace{1cm}+\widetilde{\Theta}'(k)\Phi_A'(k, k+\bar{T})(I-J)^2\Phi_A(k, k+\bar{T})\widetilde{\Theta}(k)\\&=\widetilde{\Theta}'(k)J\widetilde{\Theta}(k)+\widetilde{\Theta}'(k)Q_\alpha(k)\widetilde{\Theta}(k)\\&\hspace{1cm}+(1-m)^2\widetilde{\Theta}'(k)\tilde{P}'(k)(I-J)\tilde{P}(k)\widetilde{\Theta}(k)\\&\leq \widetilde{\Theta}'(k)J\widetilde{\Theta}(k)+\widetilde{\Theta}'(k)Q_\alpha(k)\widetilde{\Theta}(k)\\&\hspace{1cm}+(1-m)^2\widetilde{\Theta}'(k)(I-J)\widetilde{\Theta}(k)\\&=\widetilde{\Theta}'(k)\widetilde{\Theta}(k)+\widetilde{\Theta}'(k)Q_\alpha(k)\widetilde{\Theta}(k)\\&\hspace{1cm}+(m^2-2m)\widetilde{\Theta}'(k)(I-J)\widetilde{\Theta}(k)\\&=V(k)+\widetilde{\Theta}'(k)Q_\alpha(k)\widetilde{\Theta}(k)+(m^2-2m)\tilde{V}(k),
		\end{split}
	\end{equation*}
	where the third equality holds by \eqref{eqn:P} and the inequality holds by Lemma~\ref{lmm:Ax}. Moreover, as assumed in this scenario, $V_2(k)> \frac{(1-\sqrt{nr})^2}{n+(1-\sqrt{nr})^2} V(k)$. We therefore conclude that
	\begin{equation}
		\begin{split}
			&\widetilde{\Theta}'(k)\Phi_M'(k,k+\bar{T})\Phi_M(k,k+\bar{T})\widetilde{\Theta}(k) \\&\leq \Big(1-(2m-m^2)c \Big)V(k)+\widetilde{\Theta}'(k)Q_\alpha(k)\widetilde{\Theta}(k)\\&=\Big(1-(m-m^2)c \Big)V(k)+\widetilde{\Theta}'(k)Q_\alpha(k)\widetilde{\Theta}(k)-mcV(k).
		\end{split}
	\end{equation}
	In view of \eqref{eqn:Q}, there must exist a finite time $k^*$ such that when $k>k^*$, $Q_\alpha(k)<mc$ holds. Consequently, we can obtain that
	\begin{equation}\label{eqn:scenario2}
		\begin{split}
			&\widetilde{\Theta}'(k)\Phi_M'(k,k+\bar{T})\Phi_M(k,k+\bar{T})\widetilde{\Theta}(k) \\&\leq \Big(1-(m-m^2)c \Big)V(k), \;\forall k>k^*.
		\end{split}
	\end{equation}
	Denote 
	\begin{equation}\label{eqn:gamma2}
		\gamma_2 \triangleq 
		(m-m^2)c\in(0,1).
	\end{equation}
	We therefore conclude from \eqref{eqn:errmean} that
	\begin{equation}\label{eqn:case2}
		\begin{split}
			\mathbb{E}[V(k+\bar{T}+1)]\leq (1-\gamma_2)\mathbb{E}[V(k)]+ C\alpha^2(k).
		\end{split}
	\end{equation}
	
	We finally combine the two scenarios discussed so far. Let
	\begin{equation}
		\begin{split}
			&\gamma(k) \triangleq \min(\gamma_1(k),\gamma_2)\in(0,1).
		\end{split}
	\end{equation}
	By \eqref{eqn:case1} and \eqref{eqn:case2}, the dynamics of $\mathbb{E}[V(k)]$ is bounded as
	\begin{equation}
		\begin{split}
			\mathbb{E}[V(k+\bar{T}+1)]\leq (1-\gamma(k))\mathbb{E}[V(k)]+ C\alpha^2(k).
		\end{split}
	\end{equation}
	This is in the form of \eqref{eqn:lemma6} and thus we are ready to apply Lemma~\ref{lmm:converge}.

	In view of \eqref{eqn:zeta2} and \eqref{eqn:gamma1}, we know that $\gamma_1(k)$ increases with $\alpha(k)$.
	Moreover, notice that $\{\alpha(k)\}$ is monotonically non-increasing and $\gamma_2$ is a constant. Hence, there must exist a finite time $\tilde{k}$ such that 
	\begin{equation}\label{eqn:gamma}
		\gamma(k) =\begin{cases}
			\gamma_2, &\text{ if }k< \tilde{k},\\\gamma_1(k), &\text{ otherwise}.
		\end{cases}
	\end{equation}
	Thus, we know that
	\begin{equation*}
		\begin{split}
			\sum_{k=0}^\infty \gamma(k) &= \sum_{k=0}^{\tilde{k}-1}\gamma_2+\sum_{k=\tilde{k}}^{\infty}\gamma_1(k)
			\\&\geq r(1-c)\sum_{k=\tilde{k}}^{\infty}\Big(\frac{2\alpha(k)\underline{\omega}}{\bar\mu+\underline{\omega}}-\frac{\alpha^2(k)\underline{\omega}^2}{(\bar\mu+\underline{\omega})^2}\Big)= \infty,
		\end{split}
	\end{equation*}
	where the last equality holds by \eqref{eqn:alpha}.
	Moreover, one can also verify that
	\begin{equation}
		\begin{split}
			\lim_{k \to \infty} \frac{C\alpha^2(k)}{\gamma(k)} &= \lim_{k \to \infty} \frac{C\alpha^2(k)}{\gamma_1(k)} \\&=\frac{C(\bar\mu+\underline{\omega})}{r(1-c)\underline{\omega}}\lim_{k \to \infty}\frac{\alpha^2(k)}{\alpha(k)}\\&\leq\frac{C(\bar\mu+\underline{\omega})(\bar{T}+1)}{r(1-c)\underline{\omega}}\lim_{k \to \infty}\alpha(k)=0.
		\end{split}
	\end{equation}
	Therefore, all conditions in Lemma~\ref{lmm:converge} are verified. It thus follows
	$
	\lim_{k \to \infty} \mathbb{E}[\widetilde{\Theta}'(k)\widetilde{\Theta}(k)]=\lim_{k \to \infty} \mathbb{E}[V(k)] =0.
	$\hfill$\square$
\end{pf*}

\begin{remark}\label{rmk:twopart}
	\rm We briefly summarize intuitions in constructing the proof of Theorem~\ref{thm:converge}. The structure of the Lyapunov function $V(k)$ in \eqref{eqn:Lyapunov} is of importance. 
	It is decomposed into two parts $V_1(k)$ and $V_2(k)$, which represent different aspects of the error as follows. By \eqref{eqn:Jdef} and \eqref{eqn:J}, we can rewrite $V_1(k)$ in \eqref{eqn:V} as
	\begin{equation*}
	V_1(k) = (J\widetilde{\Theta}(k))^\prime (J \widetilde{\Theta}(k))=n(\nu(k))^2,
	\end{equation*}
	where
	$
	\nu(k)\triangleq\frac{1}{n} \sum_{i=1}^{n} \widetilde{\theta}_{i}(k).
	$
	That is, $\nu(k)$ is the average of the local estimation errors from all sensors at time $k$. Similarly, $V_2(k)$ can be expressed as
	\begin{equation*}
	V_2(k)=((I-J)\widetilde{\Theta}(k))^\prime ((I-J) \widetilde{\Theta}(k)) = \sum_{i=1}^{n}(\widetilde{\theta}_i(k)-\nu(k))^2,
	\end{equation*}
	which is the variance of the local estimation errors. For simplicity, let us denote 
	$
	\sigma(k)\triangleq\sum_{i=1}^{n}(\widetilde{\theta}_i(k)-\nu(k))^2.
	$
	Therefore, the Lyapunov function \eqref{eqn:Lyapunov} is equivalent to
	\begin{equation*}
	V(k) = n(\nu(k))^2+\sigma(k).
	\end{equation*}
It is evident that $V(k)$ converges to zero if and only if both the average and variance of local estimation errors, i.e., $\nu(k)$ and $\sigma(k)$, go to zero. Specifically, when certain sensors are sufficiently excited, they tend to approach the true parameter, resulting in a reduction in the average estimation error $\nu(k)$ during the adaptation step. Additionally, the variance among these sensor estimates, namely $\sigma(k)$, decreases as local estimates become more evenly distributed among sensors during the combination step. It is worth noting that this is aligned with the two scenarios mentioned in the above proof.

%		Intuitively, the estimation performance is guaranteed by both the adaptation and combination steps in Algorithm~\ref{alg:CTA}. Specifically, the adaptation step ensures that the excited sensors move towards the true parameter. On the other hand, in the combination step, each sensor performs a consensus algorithm to fairly mix the local information in its neighborhood, in turn reducing the variance among all sensors. As will be seen later, these respectively correspond to Scenarios I and II in our proof, where certain properties of matrices $G(k)$ and $A(k)$ play a key role.
	
\end{remark}

In many existing works on distributed parameter estimation, the authors study the stability of local estimation error by initially examining the stability of deterministic systems, and then extending the results to stochastic systems using the stochastic internal-external stability theory (\hspace{1pt}\cite{xie2015stability,yu2019robust,xie2018analysis,guo1994stability}). By doing so, they show the local estimate of each sensor converges to a neighborhood of the true parameter in mean square.
	
In contrast, our approach presents a novel proof framework in stochastic settings, by decomposing the Lyapunov function into two parts, which respectively correspond to the average and variance of local estimation errors. We show that the first part decreases solely in the combination step, while the second part decreases solely in the adaptation step, as discussed in Remark~\ref{rmk:twopart}. This decoupled structure enables us to independently design and analyze the combination and adaptation steps. For instance, while the adaptation step is designed through DREM in this paper, it can be readily replaced with other models. To our knowledge, this is the first time this decoupling result is proposed within the domain of distributed estimation. Moreover, by using stochastic approximation techniques and probability limit theory, we obtain a more precise result indicating that the local estimate of each sensor converges to the actual parameter itself, rather than merely approaching a neighborhood around it.

Finally, as compared to existing works in \cite{chen2013distributed,xie2018analysis,xie2020convergence,matveev2021diffusion,abdolee2014diffusion,yan2023distributed,tu2012diffusion}, our algorithm works under milder requirements on network topology and excitation condition, where it is possible that no individual sensor is persistently excited, and the network may experience disconnections (see Table~\ref{tab:compare}).

\subsection{Extension to ATC diffusion-based algorithm}\label{sec:ATC}
It is worth noting that the CTA-type solution in Algorithm~\ref{alg:CTA} can be slightly modified to obtain an ATC diffusion-based estimation scheme. Such a version is outlined in Algorithm~\ref{alg:ATC}. 

\begin{algorithm}[h!] 
	\textbf{for} $\ell\in\{1,2,...,d\}$ \textbf{do}
	\qquad \begin{itemize}
		\item[1):] Adapt the $\ell$-th entry of local estimate as
		
		\textit{(Adaptation)  }
		\begin{equation}
			\begin{split}
				&\bar{\theta}^{\ell,A}_{i}(k+1)=\widehat{\theta}_i^{\ell,A} (k)
				\\&\;+\frac{\alpha(k)\delta_{i}(k)}{\mu_{i}+\left(\delta_{i}(k)\right)^{2}}\left(\overline{y}^{\ell}_{i}(k)-\delta_{i}(k) \widehat{\theta}_i^{\ell,A} (k)\right),
			\end{split}
		\end{equation}
		where $\mu_i>0$ and $\alpha(k)$ is the stepsize to be designed later.
		
		\item[2):] Combine the local estimates in neighborhood as
		
		\textit{(Combination)  } 
		\begin{equation}
			\begin{split}
				&\widehat{\theta}_i^{\ell,A} (k+1) =
				a_{ii}(k)\bar{\theta}^{\ell,A}_{i}(k+1)\\&\;+\sum_{j\in\mathcal{N}^+_i(k)}a_{ij}(k)\bar{\theta}^{\ell,A}_{j}(k+1),
			\end{split}
		\end{equation}
		where $a_{ij}(k)$ is the $(i,j)$-th entry of the weighted adjacency matrix $A(k)$. 
	\end{itemize}
	\textbf{end for}
	\caption{ATC diffusion-based estimation algorithm under Gaussian observation noises}
	\label{alg:ATC}
\end{algorithm}

As shown in \cite{xie2020convergence} and \cite{matveev2021diffusion}, the performance of ATC algorithms can be analyzed in a similar way as their CTA counterparts.
Therefore, we will present the following result regarding the performance of Algorithm~\ref{alg:ATC} without giving the detailed proof:

\begin{corollary}\label{col:converge}
	Consider the network of sensors satisfying the stochastic linear regression model \eqref{eqn:LRE}.
	Suppose that Assumption~\ref{assup:assumptions} holds. Then, by performing Algorithm~\ref{alg:ATC}, it holds for any $\ell \in\{1,\cdots,d\}$ that
	\begin{align}
		\lim\limits_{k\to \infty} &\mathbb{E}[(\widetilde{\Theta}^{\ell,A}(k))'\widetilde{\Theta}^{\ell,A}(k)]=0,
	\end{align}
	where $\widetilde{\Theta}^{\ell,A}(k)\triangleq \col(\widetilde{\theta}^{\ell,A}_{1}(k), \ldots, \widetilde{\theta}^{\ell,A}_{n}(k))$ and $\widetilde{\theta}^{\ell,A}_{i}(k)\triangleq \widehat{\theta}^{\ell,A}_{i}(k)-\theta^{\ell,A}_{i}(k) ,\forall i\in\mathcal{V}.$That is, each sensor infers the true parameter $\theta$ in the mean square sense. 
\end{corollary}

\section{Numerical Examples}\label{sec:simulation}

In this section, we will present some numerical examples to demonstrate the theoretical results established in the previous sections and also to compare our algorithm with those in the literature. 

\subsection{Performance of Algorithm~\ref{alg:CTA}}
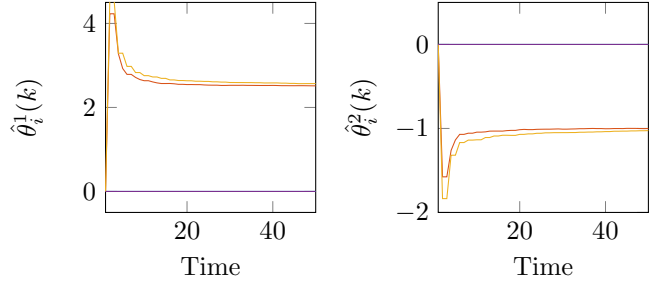
\begin{figure}
	%\centering
	\begin{minipage}[b]{0.1\textwidth}
		\input{tikz/est_compare.tikz}
	\end{minipage}\hspace{73pt}
	\begin{minipage}[b]{0.1\textwidth}
		\input{tikz/est2_compare.tikz}
	\end{minipage}
	\caption{Average of the local estimate $\hat{\theta}_i(k)$ of each sensor by performing individually in $1000$-run Monte Carlo trials.}
	\label{fig:compare}
\end{figure}

\begin{figure}[!htbp]
	\centering
	\includegraphics[width=0.4\textwidth]{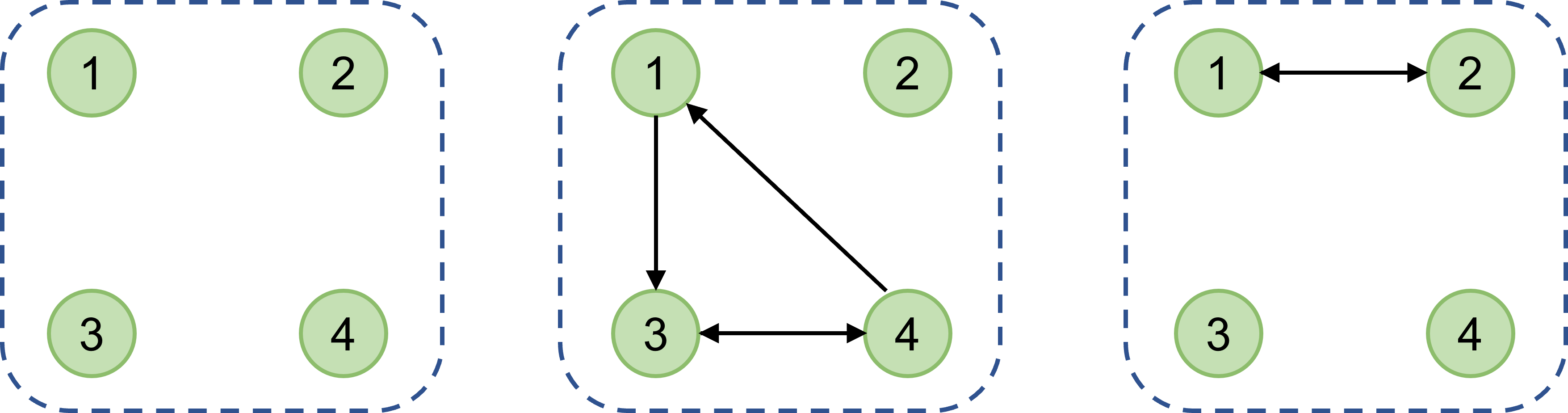}
	\caption{The communication network of sensors.}
	\label{fig:network}
\end{figure}

Let us consider the network of $4$ sensors, which aim to estimate a $2$-dimensional parameter $\theta$ over the network. Assume that the true parameter is 
$
	\theta = [2.5 \quad  -1]^\prime.
$
Moreover, the regressor $\phi_i(k)$ of each sensor is designed as
\begin{equation}
	\begin{split}
		\phi_1(k)&=
		[1 \quad 0 ]^\prime,
		\;\phi_2(k)=[
		a(k) \quad 1
		]^\prime,\\\phi_3(k)&=[
		1\quad b(k)
		]^\prime,\;\phi_4(k)=[1 \quad 1 ]^\prime,
	\end{split}
\end{equation}
where 
$
	a(k)=a(k-1)+\cos\left(\frac{k\pi}{4}\right), b(k)=b(k-1)+\sin\left(\frac{k\pi}{2}\right),
$
with $a(0)=1$ and $b(0)=2$. 

We first test the estimation performance of sensors when no communication occurs among them. From Fig.~\ref{fig:compare}, it is obvious that sensors $1$ and $4$ cannot converge to the true value. This is because they do
not satisfy the conventional PE condition and their scalar regressors $\overline{\delta}_i(k)$ remain as $0$ throughout the execution. However, we can easily check that the sensors collectively satisfy the cooperative PE condition as in \eqref{eqn:PE}. Therefore, Algorithm~\ref{alg:CTA} is applicable.

To see this, suppose that the communication topology of sensors switches among the subgraphs in Fig.~\ref{fig:network}, where the adjacency matrix is set as
\begin{equation}
	A(k) = \begin{cases}
		A_1,\text{ if } k \text{ mod } 3 = 1,\\
		A_2,\text{ if } k\text{ mod } 3 = 2,\\
		A_3,\text{ if } k\text{ mod } 3 = 0,
	\end{cases}
\end{equation}
with
\begin{equation}
	\begin{split}
		A_1 &= I,\; A_2 = \begin{bmatrix}
			0.4 & 0 & 0 & 0.6\\
			0 & 1 & 0 & 0\\
			0.6 & 0 & 0.2 & 0.2\\
			0 & 0 & 0.8 & 0.2
		\end{bmatrix}, A_3 = \begin{bmatrix}
			0.3 & 0.7 & 0 & 0\\
			0.7 & 0.3 & 0 & 0\\
			0 & 0 & 1 & 0\\
			0 & 0 & 0 & 1
		\end{bmatrix}.
	\end{split}
\end{equation}
Notice that each of the subgraphs is disconnected. However, the union of them over every
time interval with a length of at least $3$ is strongly connected. 
In Algorithm~\ref{alg:CTA}, let the stepsize be $\alpha(k) = 1.8/k$, which meets the condition \eqref{eqn:alpha}.
For each sensor $i\in\mathcal{V}$, we set its initial estimate as $[0\quad 0]^\prime$. Moreover, other parameters are chosen as
$
		\mu_i = 0.1i $ and $R_i = i\cdot I, \forall i.
$

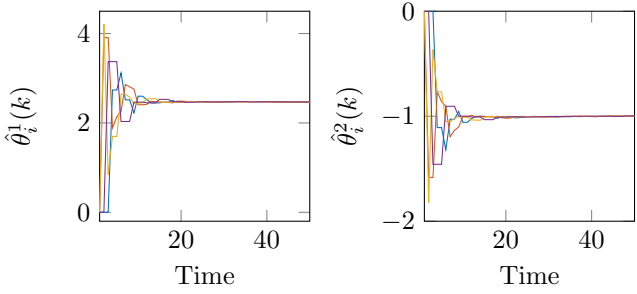
\begin{figure}
	%\centering
	\begin{subfigure}[b]{0.1\textwidth}
		\input{tikz/est.tikz}
	\end{subfigure}\hspace{70pt}
	\begin{subfigure}[b]{0.1\textwidth}
		\input{tikz/est2.tikz}
	\end{subfigure}
	\caption{Average of the local estimate $\hat{\theta}_i(k)$ of each sensor by performing Algorithm~\ref{alg:CTA} in $1000$-run Monte Carlo trials.}
	\label{fig:est}
\end{figure}

In this example, we run a Monte Carlo for $1000$ times with the same initial
states and parameters. The performance of Algorithm~\ref{alg:CTA} is demonstrated in Figs.~\ref{fig:est}. %and \ref{fig:err}
From the figure, we can see that each sensor consistently infers the true parameter, as expected from Theorem~\ref{thm:converge}.

%\begin{figure}[!htbp]
%	\centering
%	\input{tikz/error.tikz}
%	\caption{Average of the Euclidean norm of estimation error of each sensor by performing Algorithm~\ref{alg:CTA} in $1000$-run Monte Carlo trials.}
%	\label{fig:err}
%\end{figure}

\subsection{Comparison with different algorithms}
In the second example, we consider a network of $n=30$ sensors. The sensors identify the unknown parameter of the regression model in \eqref{eqn:LRE} with $d=5$, where each entry of $\theta$ is randomly drawn from a Gaussian distribution $\mathcal{N}(0,1)$.
The sensors communicate over a time-invariant network which is in a ring structure. Moreover, only sensors in $i\in\mathcal{V}'=\{1,2,\ldots,6\}$ will be sufficiently excited in turn with
\begin{equation}
	\phi_{i}(k) = \begin{cases}
		e_1, &\text{if } k \mod 30 = i,\\
		e_2, &\text{if } k \mod 30 = 6+i,\\
		e_3, &\text{if } k \mod 30 = 12+i,\\
		e_4, &\text{if } k \mod 30 = 13+i,\\
		e_5, &\text{if } k \mod 30 = 14+i,\\
		0, &\text{otherwise}, 
	\end{cases}
\end{equation}
where $e_p$ is the $p$-th canonical vector in the $5$-dimensional space for $p\in\{1,\ldots,5\}$.
On the other hand, sensors $j \in \mathcal{V}\backslash \mathcal{V}'$ remain not excited with $\phi_j(k)=0$ for any $k$. It can be verified that the entire network satisfies the cooperative PE condition in \eqref{eqn:PE}. We assume $R_i =I,\mu_i=1$ for each $i\in\mathcal{V}$, and initialize each sensor's local estimate randomly. 

\begin{figure}[!htbp]
	\centering
	\input{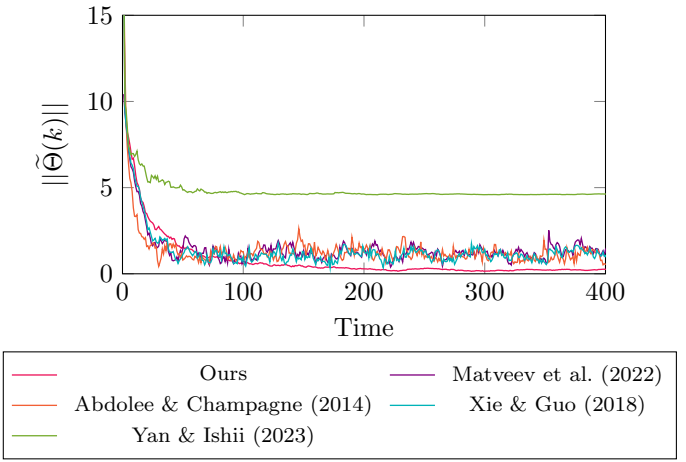}
	\caption{Comparison with different distributed parameter estimation algorithms in time-invariant communication topology.}
	\label{fig:comparison2}
\end{figure}

We compare our proposed Algorithm 1 with those in Table~\ref{tab:compare}
from \cite{abdolee2014diffusion,matveev2021diffusion,xie2018analysis,yi2022conditions}. 
Since some of them are designed only for fixed communication graphs,
we first study their performance using
the above ring network with time-invariant topology. 
In Fig.~\ref{fig:comparison2}, the results of the total estimation error 
are presented. It is worth noticing that the algorithms proposed in \cite{matveev2021diffusion,abdolee2014diffusion,xie2018analysis} demonstrate higher convergence speeds at the beginning stage. This is because constant stepsizes are used therein. Instead, we use vanishing stepsizes, which decrease the convergence rate. However, as observed from the figure, they enable the convergence of our algorithm in the presence of stochastic noises. In comparison, other works require more stringent conditions on the excitation signals (as shown in Table~\ref{tab:compare}), which cannot be met in this example, leading to worse asymptotic estimation performance than that of our method. Moreover, only stability of estimation error can be established in these works.

\begin{figure}[!htbp]
	\centering
	\input{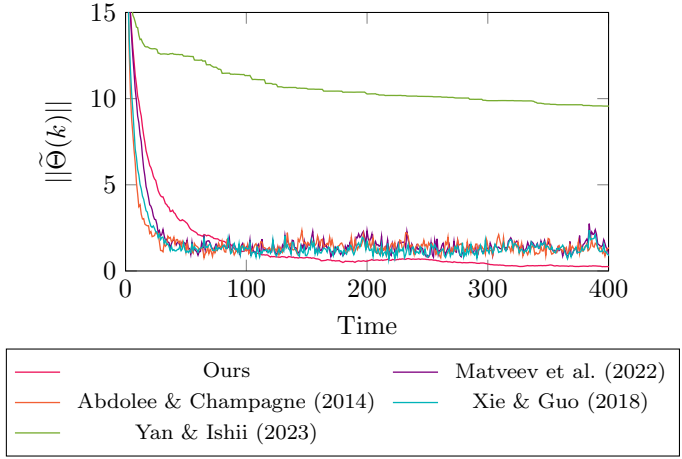}
	\caption{Comparison with different distributed parameter estimation algorithms in time-varying communication topology.}
	\label{fig:comparison}
\end{figure}

We further test the estimation performance in a time-varying communication graph. We assume that at any time, only one edge in the ring structure is activated, and the activation is performed in a periodic manner allowing communication over only $1$ edge at each time. This results in a topology that is jointly strongly connected. As illustrated in Fig.~\ref{fig:comparison}, the convergence rates of all algorithms are inevitably reduced due to the time-varying topology. Furthermore, the estimation error of \cite{yan2023distributed} is highly increased, since the Local-PE condition required therein is severely violated in this time-varying graph. However, it is clear that our algorithm is capable to filter the effects of noises, and the estimation error continuously decreases to $0$ over time.

\subsection{Temperature monitoring example}
\begin{figure}[!htbp]
	\centering
	\input{code2/temp_topo.tikz}
	\caption{The positions of targets in the grid.}
	\label{fig:temp_topo}
\end{figure}
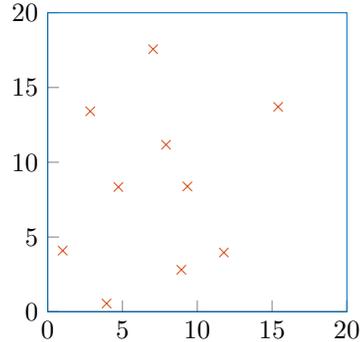

Lastly, similar to \cite{kar2012distributed}, we formulate the problem of temperature monitoring by using \eqref{eqn:LRE}. Specifically, let us consider a network of $n = 100$ sensors, deployed to monitor the temperature of $10$ distinct targets ($d = 10)$ within a region represented by a $20 \times 20$ grid, see Fig.~\ref{fig:temp_topo}. The temperature of the $j$-th target is denoted as $\theta_j$. 
	Each sensor $i$ records a noisy temperature reading at its current location. This measurement is a linear combination of the temperatures of all the targets, with weighting factors determined by the distances between the sensor and the targets. This relationship can be expressed as:
	\begin{equation}
	\begin{split}
	y_i(k) &= \sum_{j=1}^d \frac{\beta}{(d_{i,j}(k))^3} \theta_{j} + w_i(k)\\&= \theta' \phi_{i}(k) + w_i(k),
	\end{split}
	\end{equation}
	where $\theta = [\theta_1, \cdots, \theta_{10}]'$, $\phi_{i}(k) = \left[\frac{\beta}{d_{i,1}(k)}, \cdots, \frac{\beta}{d_{i,10}(k)}\right]'$, and $d_{i,j}(k) = ||p_i(k) - p_j||$ represents the distance between sensor $i$ and target $j$ at time $k$. Here, $p_i(k)$ and $p_j$ are the positions of sensor $i$ and target $j$, respectively. In this example, we choose $\beta=10$.
	
	Sensors in the set $\mathcal{V}'=\{1,2,\ldots,\frac{n}{2}\}$ are mobile within the region, while sensors in the set $\mathcal{V}\backslash \mathcal{V}'$ remain static. Consequently, sensors in $\mathcal{V}\backslash \mathcal{V}'$ do not meet the PE condition. Nevertheless, it can be confirmed that the entire network complies with the cooperative PE condition as defined in \eqref{eqn:PE}.
	Each sensor has a communication radius of $\tilde{r}$, meaning that it can only communicate with other sensors located within the distance of $\tilde{r}$. 
	
\begin{figure}[!htbp]
	\centering
	\input{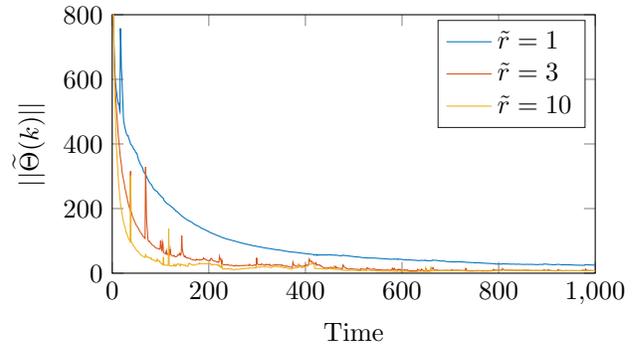}
	\caption{Local estimation error by performing Algorithm~\ref{alg:CTA} with different communication radius.}
	\label{fig:performance_r}
\end{figure}

We test different cases with $\tilde{r} = \{1,3, 10\}$. The performance is shown in Fig.~\ref{fig:performance_r}. Clearly, our algorithm helps to reduce the estimation error. Moreover, a larger $\tilde{r}$ yields a smaller finite-time estimation error and a larger convergence rate. This is because that a denser communication topology is adopted.

%\begin{figure}[!htbp]
%	\centering
%	\input{code2/performance.tikz}
%	\caption{Performance of Algorithm~\ref{alg:CTA}.}
%	\label{fig:performance}
%\end{figure}

%\begin{figure}[!htbp]
%	\centering
%	\input{code2/temp_100.tikz}
%	\caption{Performance of Algorithm~\ref{alg:CTA} with different number of sensors.}
%	\label{fig:performance_n}
%\end{figure}

%\begin{figure}[!htbp]
%	\centering
%	\input{code2/r20.tikz}
%	\caption{Performance of Algorithm~\ref{alg:CTA} with different number of sensors.}
%	\label{fig:performance_r}
%\end{figure}

%%%%%%%%%%%%%%%%%%%%%%%%%%%%%%%%%%%%%%%%%%%%%%%%%%%%%%%%%%%%%%%%%%%%%%
\section{Conclusion}\label{sec:conclude}

This paper has studied the problem of distributed parameter estimation in sensor networks, where measurements of sensors are subject to Gaussian stochastic noises. By leveraging the stochastic DREM algorithm, both CTA and ATC diffusion-based estimators have been proposed, which  guarantee that each sensor estimates the true parameter under mild conditions on excitation condition and network topology. We have introduced a novel proof framework characterized by a decoupled structure. This framework relates the average and variance of local estimation errors respectively to the combination and adaptation steps in the proposed algorithms. By means of numerical examples, we have shown that our algorithm has superior convergence properties over other distributed algorithms in the literature.

%Therefore, the sensors can track a dynamic process from noisy measurements by working cooperatively with each other, even when none of them can do so individually. Moreover, since no stringent requirements such as statistical independency and stationarity are imposed, the proposed algorithms are applicable even in stochastic feedback systems. We have demonstrated by means of numerical examples that our proposed algorithm has superior convergence properties over other distributed algorithms from recent works in the literature. %For future works, we plan to consider more challenging problems where the parameter to be estimated could be changing dynamically.

\appendix
\section{Proof of Proposition~\ref{prop:PE}}\label{app:A}
%To prove this proposition, we need the following lemma:
%\begin{lemma}\label{lmm:lemma1}
%	Consider any matrix $A\in\mathbb{R}^{d\times d}$ such that $A\geq 0$. The following statements are equivalent:
%	\begin{enumerate}
%		\item[(1)]  $A\geq \nu I$;
%		\item[(2)]
%		$
%		\sigma_i \geq \nu, \;\forall i,
%		$
%	\end{enumerate}
%	where $\{\sigma_i\}$ are the eigenvalues of $A$ and $\nu >0$.
%\end{lemma}
%\begin{pf*}{Proof.}
%	Since $A\geq 0$, there exists $Q\in\mathbb{R}^{d\times d}$ such that $A=Q'\Lambda Q$, where $\Sigma\triangleq \diag\{\sigma_1,\cdots,\sigma_d\}$. Therefore, we can diagonalize $A- \nu I$ as 
%	$
%	A- \nu I = Q'(\Sigma-\nu I) Q = Q'\diag\{\sigma_1-\nu,\cdots,\sigma_d-\nu\}Q.
%	$
%	The proof is then straightforward by noticing that $A- \nu I \geq 0$ holds, if and only if  $\diag\{\sigma_1-\nu,\cdots,\sigma_d-\nu\}\geq0$.\hfill$\square$
%\end{pf*}
%
%Now we are ready to prove Proposition~1:

As discussed previously, the cooperative PE condition \eqref{eqn:PE} is equivalent to \eqref{eqn:PE2}.
	By the definition of $\delta_i(t)$, we rewrite \eqref{eqn:PE2} as
	\begin{equation}\label{eqn:R5}
	\begin{split}
	&\det \left(\begin{bmatrix}
	\phi_i(t) & \cdots &\phi_i(t-d+1)
	\end{bmatrix}\begin{bmatrix}
	\phi'_i(t) \\ \vdots \\\phi'_i(t-d+1)
	\end{bmatrix}\right) \\&= \det[\phi_i(t)\phi'_i(t)+\cdots+\phi_i(t-d+1)\phi'_i(t-d+1)]  \\&=\det(\mathcal{A}(t))\geq \underline{\omega},
	\end{split}
	\end{equation}
	where $\mathcal{A}(t)\triangleq \sum_{\tau=t-d+1}^{t} \phi_i(\tau)\phi'_i(\tau)\in\mathbb{R}^{d\times d}$. 
	
	On the other hand, \eqref{eqn:PE_2} implies 
	\begin{equation}\label{eqn:R6}
	\mathcal{A}(t)\geq \omega_2 I. 
	\end{equation} 
	We shall show the equivalence between \eqref{eqn:R5} and \eqref{eqn:R6}. For simplicity, let us denote the eigenvalues of $\mathcal{A}(t)$ as $\{\lambda_j(t)\}$. Since $\mathcal{A}(t)\geq 0$, $\lambda_j(t)\geq 0$ for any $j$.

	\eqref{eqn:R5}$\Rightarrow$\eqref{eqn:R6}: As the signal $\phi_i(t)$ is bounded at any $t$ (Assumption~\ref{assup:assumptions}), there exists $\rho>0$ such that each entry of $\mathcal{A}(t)$ is upper bounded by $\rho$. For any $\lambda_j(t)$, let us denote its corresponding eigenvector as $q_j(t)\neq 0$. We thus have
	\begin{equation*}
	\lambda_j ||q_j(t)||_\infty = ||\mathcal{A} q_j(t)||_\infty\leq d\rho||q_j(t)||_\infty.  
	\end{equation*}
	We therefore conclude $\lambda_j(t) \leq d\rho.$ 
	
	On the other hand, by \eqref{eqn:R5}, $\det(\mathcal{A}(t))=\prod_{j=1}^d \lambda_j(t) \geq \underline{\omega}$. We have $$\lambda_j(t) \geq \frac{\underline{\omega}}{\lambda_1(t)\cdots \lambda_{j-1}(t)\lambda_{j+1}(t)\lambda_{d}(t)} \geq \frac{\underline{\omega}}{(d\rho)^{d-1}}, \;\forall j.$$
	It is thus concluded $\mathcal{A}(t)\geq \frac{\underline{\omega}}{(d\rho)^{d-1}}I$. By setting $\omega_2=\frac{\underline{\omega}}{(d\rho)^{d-1}}$, we arrive at \eqref{eqn:R6}.
	
	\eqref{eqn:R6}$\Rightarrow$\eqref{eqn:R5}: From \eqref{eqn:R6}, it is not difficult to verify $\lambda_j(t)\geq \omega_2$ holds for any $j$. Then, we have $\det(\mathcal{A}(t))=\prod_{j=1}^d \lambda_j(t) \geq \omega_2^d.$ Setting $\underline{\omega} = \omega_2^d$, we finish the proof.

\bibliographystyle{agsm} 
\bibliography{reference}

\end{document}

%% file: tikz/est_compare.tikz
\definecolor{mycolor1}{rgb}{0.00000,0.44700,0.74100}%
\definecolor{mycolor2}{rgb}{0.85000,0.32500,0.09800}%
\definecolor{mycolor3}{rgb}{0.92900,0.69400,0.12500}%
\definecolor{mycolor4}{rgb}{0.49400,0.18400,0.55600}%
\begin{tikzpicture}

\begin{axis}[%
width=1.1in,
height=1.1in,
at={(0.in,0.in)},
scale only axis,
xmin=1,
xmax=50,
xlabel={{Time}},
ymin=-0.5,
ymax=4.5,
ylabel={{$\hat{\theta}^1_i(k)$}},
ytick={0,2,4},
yticklabels={$\;\;\;0$, $\;2$,$\;4$},
axis background/.style={fill=white}
]
\addplot [color=mycolor1, forget plot]
  table[row sep=crcr]{%
1	0\\
2	0\\
3	0\\
4	0\\
5	0\\
6	0\\
7	0\\
8	0\\
9	0\\
10	0\\
11	0\\
12	0\\
13	0\\
14	0\\
15	0\\
16	0\\
17	0\\
18	0\\
19	0\\
20	0\\
21	0\\
22	0\\
23	0\\
24	0\\
25	0\\
26	0\\
27	0\\
28	0\\
29	0\\
30	0\\
31	0\\
32	0\\
33	0\\
34	0\\
35	0\\
36	0\\
37	0\\
38	0\\
39	0\\
40	0\\
41	0\\
42	0\\
43	0\\
44	0\\
45	0\\
46	0\\
47	0\\
48	0\\
49	0\\
50	0\\
};
\addplot [color=mycolor2, forget plot]
  table[row sep=crcr]{%
1	0\\
2	4.23021409109044\\
3	4.23021409109044\\
4	3.27579430647541\\
5	2.92715826877365\\
6	2.78691814645455\\
7	2.78691814645455\\
8	2.71489496446064\\
9	2.66433056584895\\
10	2.6363649535684\\
11	2.6363649535684\\
12	2.60739882189477\\
13	2.58458544177528\\
14	2.56939330687964\\
15	2.56939330687964\\
16	2.57047946361076\\
17	2.56317136322812\\
18	2.55021072357381\\
19	2.55021072357381\\
20	2.54446820007935\\
21	2.54537334819143\\
22	2.54315322064391\\
23	2.54315322064391\\
24	2.53546945873494\\
25	2.53131704010487\\
26	2.52995533316722\\
27	2.52995533316722\\
28	2.52353711230198\\
29	2.52395713337755\\
30	2.52689568822798\\
31	2.52689568822798\\
32	2.52637565441941\\
33	2.52801869289227\\
34	2.52507633581525\\
35	2.52507633581525\\
36	2.52353283188647\\
37	2.52274142304266\\
38	2.52280756029351\\
39	2.52280756029351\\
40	2.52435368073184\\
41	2.52203252284537\\
42	2.51836693707808\\
43	2.51836693707808\\
44	2.51739018537165\\
45	2.51720695816866\\
46	2.51792882782113\\
47	2.51792882782113\\
48	2.51651908358977\\
49	2.5150040336549\\
50	2.51416944116938\\
};
\addplot [color=mycolor3, forget plot]
  table[row sep=crcr]{%
1	0\\
2	4.5460364714636\\
3	4.5460364714636\\
4	3.29109398325447\\
5	3.29109398325447\\
6	2.9772084511949\\
7	2.9772084511949\\
8	2.83051902113032\\
9	2.83051902113032\\
10	2.75678112252769\\
11	2.75678112252769\\
12	2.72521235861893\\
13	2.72521235861893\\
14	2.69214253822681\\
15	2.69214253822681\\
16	2.65528841341773\\
17	2.65528841341773\\
18	2.64067988059973\\
19	2.64067988059973\\
20	2.63490161451947\\
21	2.63490161451947\\
22	2.62431166168641\\
23	2.62431166168641\\
24	2.61949962487749\\
25	2.61949962487749\\
26	2.6100028176672\\
27	2.6100028176672\\
28	2.6031176223846\\
29	2.6031176223846\\
30	2.593423953511\\
31	2.593423953511\\
32	2.59239432605927\\
33	2.59239432605927\\
34	2.5912152434287\\
35	2.5912152434287\\
36	2.58829619837318\\
37	2.58829619837318\\
38	2.5829866622706\\
39	2.5829866622706\\
40	2.58386492019289\\
41	2.58386492019289\\
42	2.57829843880802\\
43	2.57829843880802\\
44	2.57231715280024\\
45	2.57231715280024\\
46	2.56656511775692\\
47	2.56656511775692\\
48	2.56782426408723\\
49	2.56782426408723\\
50	2.56781033801273\\
};
\addplot [color=mycolor4, forget plot]
  table[row sep=crcr]{%
1	0\\
2	0\\
3	0\\
4	0\\
5	0\\
6	0\\
7	0\\
8	0\\
9	0\\
10	0\\
11	0\\
12	0\\
13	0\\
14	0\\
15	0\\
16	0\\
17	0\\
18	0\\
19	0\\
20	0\\
21	0\\
22	0\\
23	0\\
24	0\\
25	0\\
26	0\\
27	0\\
28	0\\
29	0\\
30	0\\
31	0\\
32	0\\
33	0\\
34	0\\
35	0\\
36	0\\
37	0\\
38	0\\
39	0\\
40	0\\
41	0\\
42	0\\
43	0\\
44	0\\
45	0\\
46	0\\
47	0\\
48	0\\
49	0\\
50	0\\
};
\end{axis}
\end{tikzpicture}%

%% file: tikz/est2_compare.tikz
\definecolor{mycolor1}{rgb}{0.00000,0.44700,0.74100}%
\definecolor{mycolor2}{rgb}{0.85000,0.32500,0.09800}%
\definecolor{mycolor3}{rgb}{0.92900,0.69400,0.12500}%
\definecolor{mycolor4}{rgb}{0.49400,0.18400,0.55600}%
\begin{tikzpicture}

\begin{axis}[%
width=1.1in,
height=1.1in,
at={(0.058in,0.481in)},
scale only axis,
xmin=1,
xmax=50,
xlabel={{Time}},
ymin=-2,
ymax=0.5,
ylabel={{$\hat{\theta}^2_i(k)$}},
ytick={-2,-1,0},
yticklabels={$-2$,$-1$,$0$},
axis background/.style={fill=white}
]
\addplot [color=mycolor1, forget plot]
  table[row sep=crcr]{%
1	0\\
2	0\\
3	0\\
4	0\\
5	0\\
6	0\\
7	0\\
8	0\\
9	0\\
10	0\\
11	0\\
12	0\\
13	0\\
14	0\\
15	0\\
16	0\\
17	0\\
18	0\\
19	0\\
20	0\\
21	0\\
22	0\\
23	0\\
24	0\\
25	0\\
26	0\\
27	0\\
28	0\\
29	0\\
30	0\\
31	0\\
32	0\\
33	0\\
34	0\\
35	0\\
36	0\\
37	0\\
38	0\\
39	0\\
40	0\\
41	0\\
42	0\\
43	0\\
44	0\\
45	0\\
46	0\\
47	0\\
48	0\\
49	0\\
50	0\\
};
\addplot [color=mycolor2, forget plot]
  table[row sep=crcr]{%
1	0\\
2	-1.57616528631316\\
3	-1.57616528631316\\
4	-1.26135189055125\\
5	-1.13725763143278\\
6	-1.07073752920777\\
7	-1.07073752920777\\
8	-1.0566904116295\\
9	-1.05694055664842\\
10	-1.04407562109273\\
11	-1.04407562109273\\
12	-1.04085962796449\\
13	-1.03201189037685\\
14	-1.03150891332946\\
15	-1.03150891332946\\
16	-1.03191455858847\\
17	-1.02584345884361\\
18	-1.02321137822223\\
19	-1.02321137822223\\
20	-1.0149183926488\\
21	-1.01145146267096\\
22	-1.01393488824711\\
23	-1.01393488824711\\
24	-1.01068533257037\\
25	-1.00961183662195\\
26	-1.00891127937646\\
27	-1.00891127937646\\
28	-1.00722504918742\\
29	-1.00579982423157\\
30	-1.00855967469559\\
31	-1.00855967469559\\
32	-1.00674785194073\\
33	-1.00556626301843\\
34	-1.00310893621983\\
35	-1.00310893621983\\
36	-1.00141019125701\\
37	-1.00016003005155\\
38	-1.0014871550124\\
39	-1.0014871550124\\
40	-1.00369476602771\\
41	-1.00326602367164\\
42	-1.00109555645942\\
43	-1.00109555645942\\
44	-1.00167299432765\\
45	-0.999168451183232\\
46	-0.999513277717896\\
47	-0.999513277717896\\
48	-1.00130862143858\\
49	-1.00247538622745\\
50	-1.00082900260288\\
};
\addplot [color=mycolor3, forget plot]
  table[row sep=crcr]{%
1	0\\
2	-1.83542840367895\\
3	-1.83542840367895\\
4	-1.31886577834522\\
5	-1.31886577834522\\
6	-1.16824489990887\\
7	-1.16824489990887\\
8	-1.13888307943487\\
9	-1.13888307943487\\
10	-1.13516000745961\\
11	-1.13516000745961\\
12	-1.10821827910538\\
13	-1.10821827910538\\
14	-1.09011050577191\\
15	-1.09011050577191\\
16	-1.0810375199252\\
17	-1.0810375199252\\
18	-1.08268147267296\\
19	-1.08268147267296\\
20	-1.07059839441762\\
21	-1.07059839441762\\
22	-1.064654754854\\
23	-1.064654754854\\
24	-1.05763755863411\\
25	-1.05763755863411\\
26	-1.05214492261611\\
27	-1.05214492261611\\
28	-1.05025078370509\\
29	-1.05025078370509\\
30	-1.04988580219977\\
31	-1.04988580219977\\
32	-1.0488096863633\\
33	-1.0488096863633\\
34	-1.04522294671697\\
35	-1.04522294671697\\
36	-1.04317058367167\\
37	-1.04317058367167\\
38	-1.04015569189437\\
39	-1.04015569189437\\
40	-1.0374442613543\\
41	-1.0374442613543\\
42	-1.03323531297476\\
43	-1.03323531297476\\
44	-1.03191548545238\\
45	-1.03191548545238\\
46	-1.02939037038758\\
47	-1.02939037038758\\
48	-1.02748612416833\\
49	-1.02748612416833\\
50	-1.02594452829358\\
};
\addplot [color=mycolor4, forget plot]
  table[row sep=crcr]{%
1	0\\
2	0\\
3	0\\
4	0\\
5	0\\
6	0\\
7	0\\
8	0\\
9	0\\
10	0\\
11	0\\
12	0\\
13	0\\
14	0\\
15	0\\
16	0\\
17	0\\
18	0\\
19	0\\
20	0\\
21	0\\
22	0\\
23	0\\
24	0\\
25	0\\
26	0\\
27	0\\
28	0\\
29	0\\
30	0\\
31	0\\
32	0\\
33	0\\
34	0\\
35	0\\
36	0\\
37	0\\
38	0\\
39	0\\
40	0\\
41	0\\
42	0\\
43	0\\
44	0\\
45	0\\
46	0\\
47	0\\
48	0\\
49	0\\
50	0\\
};
\end{axis}
\end{tikzpicture}%

%% file: tikz/est.tikz
\definecolor{mycolor1}{rgb}{0.00000,0.44700,0.74100}%
\definecolor{mycolor2}{rgb}{0.85000,0.32500,0.09800}%
\definecolor{mycolor3}{rgb}{0.92900,0.69400,0.12500}%
\definecolor{mycolor4}{rgb}{0.49400,0.18400,0.55600}%
\begin{tikzpicture}

\begin{axis}[%
width=1.1in,
height=1.1in,
at={(0in,0in)},
scale only axis,
xmin=1,
xmax=50,
xlabel={{Time}},
ymin=-0.2,
ymax=4.5,
ylabel={{$\hat{\theta}^1_i(k)$}},
ytick={0,2,4},
yticklabels={$\;\;\;0$, $\;2$,$\;4$},
axis background/.style={fill=white}
]
\addplot [color=mycolor1, forget plot]
  table[row sep=crcr]{%
1	0\\
2	0\\
3	0\\
4	2.73688240097793\\
5	2.73688240097793\\
6	3.11807036647615\\
7	2.51663808048513\\
8	2.51663808048513\\
9	2.22609908369877\\
10	2.59824634118981\\
11	2.59824634118981\\
12	2.51887721134449\\
13	2.44769703687445\\
14	2.44769703687445\\
15	2.44925355209614\\
16	2.46587082706459\\
17	2.46587082706459\\
18	2.5032946821136\\
19	2.46490504623302\\
20	2.46490504623302\\
21	2.47115808858681\\
22	2.48002664392192\\
23	2.48002664392192\\
24	2.46927595144576\\
25	2.47438536523034\\
26	2.47438536523034\\
27	2.47313135653786\\
28	2.46881943362255\\
29	2.46881943362255\\
30	2.47000501130061\\
31	2.47143273073158\\
32	2.47143273073158\\
33	2.47233243514374\\
34	2.4766272316594\\
35	2.4766272316594\\
36	2.47622108766054\\
37	2.47173008043798\\
38	2.47173008043798\\
39	2.47149733560683\\
40	2.47539175205563\\
41	2.47539175205563\\
42	2.47382061263517\\
43	2.46781958245103\\
44	2.46781958245103\\
45	2.4707112391196\\
46	2.46684494242881\\
47	2.46684494242881\\
48	2.46844195149239\\
49	2.47047903487062\\
50	2.47047903487062\\
};
\addplot [color=mycolor2, forget plot]
  table[row sep=crcr]{%
1	0\\
2	3.90983200139704\\
3	3.90983200139704\\
4	1.89045494878528\\
5	2.1357447559234\\
6	2.25888138648898\\
7	2.86031367248\\
8	2.80832910010816\\
9	2.75773802297169\\
10	2.40646340162698\\
11	2.40646340162698\\
12	2.41719124781587\\
13	2.48821583620785\\
14	2.47299251633678\\
15	2.47299251633678\\
16	2.45743444280737\\
17	2.45896224182909\\
18	2.44845234514134\\
19	2.48684198102192\\
20	2.487639509683\\
21	2.48382745335126\\
22	2.47367015998408\\
23	2.47367015998408\\
24	2.47657511399516\\
25	2.47403961745381\\
26	2.46697146665884\\
27	2.46697146665884\\
28	2.47002076774672\\
29	2.46858894318641\\
30	2.47204461048771\\
31	2.47061689105674\\
32	2.47275631607141\\
33	2.47846785873754\\
34	2.47379768032779\\
35	2.47379768032779\\
36	2.46980536305688\\
37	2.47774460121296\\
38	2.47706078767655\\
39	2.47706078767655\\
40	2.468908726759\\
41	2.4692715209634\\
42	2.46524771237211\\
43	2.47124874255625\\
44	2.46409021719889\\
45	2.46518795813276\\
46	2.46994463904469\\
47	2.46994463904469\\
48	2.47135207060414\\
49	2.47149815463082\\
50	2.47203458657922\\
};
\addplot [color=mycolor3, forget plot]
  table[row sep=crcr]{%
1	0\\
2	4.21524459601038\\
3	0.843048919202076\\
4	1.69745910476608\\
5	1.69745910476608\\
6	2.64380202232837\\
7	2.64380202232837\\
8	2.57435396784921\\
9	2.43133492569583\\
10	2.44637292160968\\
11	2.44637292160968\\
12	2.53796254634576\\
13	2.53796254634576\\
14	2.54773209128853\\
15	2.4682228861645\\
16	2.46209749965843\\
17	2.46209749965843\\
18	2.46754934005402\\
19	2.46754934005402\\
20	2.45880433303806\\
21	2.46576925104529\\
22	2.47484198264651\\
23	2.47484198264651\\
24	2.47730099531182\\
25	2.47730099531182\\
26	2.47042040783844\\
27	2.47317437085447\\
28	2.47346644850165\\
29	2.47346644850165\\
30	2.47147099141459\\
31	2.47147099141459\\
32	2.476704846722\\
33	2.47278705540038\\
34	2.47019013423392\\
35	2.47019013423392\\
36	2.4789621509605\\
37	2.4789621509605\\
38	2.47313093984706\\
39	2.47193267070941\\
40	2.47260546620127\\
41	2.47260546620127\\
42	2.47785067625161\\
43	2.47785067625161\\
44	2.46872352769381\\
45	2.46896425705578\\
46	2.4666878145757\\
47	2.4666878145757\\
48	2.46475399905676\\
49	2.46475399905676\\
50	2.46509078817318\\
};
\addplot [color=mycolor4, forget plot]
  table[row sep=crcr]{%
1	0\\
2	0\\
3	3.3721956768083\\
4	3.3721956768083\\
5	3.3721956768083\\
6	2.03240641917453\\
7	2.03240641917453\\
8	2.03240641917453\\
9	2.46596445811427\\
10	2.46596445811427\\
11	2.46596445811427\\
12	2.4502912289106\\
13	2.4502912289106\\
14	2.4502912289106\\
15	2.52824391881295\\
16	2.52824391881295\\
17	2.52824391881295\\
18	2.47532678348934\\
19	2.47532678348934\\
20	2.47532678348934\\
21	2.46210882312831\\
22	2.46210882312831\\
23	2.46210882312831\\
24	2.47229535074287\\
25	2.47229535074287\\
26	2.47229535074287\\
27	2.47079539641932\\
28	2.47079539641932\\
29	2.47079539641932\\
30	2.47293223808518\\
31	2.47293223808518\\
32	2.47293223808518\\
33	2.47595032499464\\
34	2.47595032499464\\
35	2.47595032499464\\
36	2.47134217238606\\
37	2.47134217238606\\
38	2.47134217238606\\
39	2.47277318635486\\
40	2.47277318635486\\
41	2.47277318635486\\
42	2.47263901023199\\
43	2.47263901023199\\
44	2.47263901023199\\
45	2.46950662420144\\
46	2.46950662420144\\
47	2.46950662420144\\
48	2.46725157650085\\
49	2.46725157650085\\
50	2.46725157650085\\
};
\end{axis}
\end{tikzpicture}%

%% file: tikz/est2.tikz
\definecolor{mycolor1}{rgb}{0.00000,0.44700,0.74100}%
\definecolor{mycolor2}{rgb}{0.85000,0.32500,0.09800}%
\definecolor{mycolor3}{rgb}{0.92900,0.69400,0.12500}%
\definecolor{mycolor4}{rgb}{0.49400,0.18400,0.55600}%
\begin{tikzpicture}

\begin{axis}[%
width=1.1in,
height=1.1in,
at={(0in,0in)},
scale only axis,
xmin=1,
xmax=50,
xlabel={{Time}},
ymin=-2,
ymax=0,
ylabel={{$\hat{\theta}^2_i(k)$}},
ytick={-2,-1,0},
yticklabels={$-2$,$-1$,$0$},
axis background/.style={fill=white}
]
\addplot [color=mycolor1, forget plot]
  table[row sep=crcr]{%
1	0\\
2	0\\
3	0\\
4	-1.10816415268579\\
5	-1.10816415268579\\
6	-1.31801932149962\\
7	-1.02871027340991\\
8	-1.02871027340991\\
9	-0.954808172830769\\
10	-1.05828870385936\\
11	-1.05828870385936\\
12	-1.02294849211935\\
13	-1.00120390968575\\
14	-1.00120390968575\\
15	-1.00497522403184\\
16	-1.00180491497577\\
17	-1.00180491497577\\
18	-1.0203756823215\\
19	-1.00639384622001\\
20	-1.00639384622001\\
21	-1.00749781971984\\
22	-1.01481220430152\\
23	-1.01481220430152\\
24	-1.00911726532911\\
25	-1.00334415019647\\
26	-1.00334415019647\\
27	-1.00604288058785\\
28	-1.00321305276651\\
29	-1.00321305276651\\
30	-1.00437740526018\\
31	-1.00585209657835\\
32	-1.00585209657835\\
33	-1.00408529088485\\
34	-1.00458625196793\\
35	-1.00458625196793\\
36	-1.00278358485431\\
37	-1.00170044498934\\
38	-1.00170044498934\\
39	-1.00096368668818\\
40	-1.0017571836937\\
41	-1.0017571836937\\
42	-1.00065141202777\\
43	-1.00076824914332\\
44	-1.00076824914332\\
45	-1.00050814523856\\
46	-0.998082335779934\\
47	-0.998082335779934\\
48	-1.00029454122395\\
49	-0.999307943913249\\
50	-0.999307943913249\\
};
\addplot [color=mycolor2, forget plot]
  table[row sep=crcr]{%
1	0\\
2	-1.58309164669398\\
3	-1.58309164669398\\
4	-0.766793015462363\\
5	-0.861784801662792\\
6	-0.90472068137147\\
7	-1.19402972946117\\
8	-1.1355599451616\\
9	-1.10263750287162\\
10	-1.00280026704621\\
11	-1.00280026704621\\
12	-0.991884802928497\\
13	-1.00503273641868\\
14	-1.0004462110946\\
15	-1.0004462110946\\
16	-1.00671668839736\\
17	-1.00873531386675\\
18	-1.00040163074795\\
19	-1.01438346684943\\
20	-1.01870362754087\\
21	-1.01794694055081\\
22	-1.0032341429772\\
23	-1.0032341429772\\
24	-1.00086995799677\\
25	-1.00503465506928\\
26	-1.00200026941451\\
27	-1.00200026941451\\
28	-1.0068812135006\\
29	-1.00840000423856\\
30	-1.00648410714328\\
31	-1.00500941582511\\
32	-1.00557868973649\\
33	-1.00480094957496\\
34	-1.00251037809495\\
35	-1.00251037809495\\
36	-1.00123624219007\\
37	-1.00405969121468\\
38	-1.00209725383893\\
39	-1.00209725383893\\
40	-1.00367882984328\\
41	-1.00542429134513\\
42	-1.00081832219284\\
43	-1.00070148507729\\
44	-0.997363201756871\\
45	-0.99704270315481\\
46	-0.999115291411888\\
47	-0.999115291411888\\
48	-0.998885116494379\\
49	-0.998895133510149\\
50	-0.999267712146314\\
};
\addplot [color=mycolor3, forget plot]
  table[row sep=crcr]{%
1	0\\
2	-1.82240345921938\\
3	-0.364480691843876\\
4	-0.767444440378633\\
5	-0.767444440378633\\
6	-1.05165329562016\\
7	-1.05165329562016\\
8	-1.02285041225467\\
9	-1.00290426765248\\
10	-1.00951470425504\\
11	-1.00951470425504\\
12	-1.03625049848836\\
13	-1.03625049848836\\
14	-1.03907288395767\\
15	-1.01003480932216\\
16	-1.002103204095\\
17	-1.002103204095\\
18	-1.00493374196503\\
19	-1.00493374196503\\
20	-1.00459234867112\\
21	-1.00640153787684\\
22	-1.00847238289075\\
23	-1.00847238289075\\
24	-1.00404851648439\\
25	-1.00404851648439\\
26	-1.00448154177443\\
27	-1.00447120530919\\
28	-1.00234586546416\\
29	-1.00234586546416\\
30	-1.0016367259165\\
31	-1.0016367259165\\
32	-1.00125040336757\\
33	-1.00434282270503\\
34	-1.00019519141462\\
35	-1.00019519141462\\
36	-0.999735756206174\\
37	-0.999735756206174\\
38	-0.999774660024577\\
39	-1.001069701896\\
40	-1.00043987056495\\
41	-1.00043987056495\\
42	-1.00073417488706\\
43	-1.00073417488706\\
44	-1.00212799540776\\
45	-1.00095349709462\\
46	-0.997521304318752\\
47	-0.997521304318752\\
48	-1.00199928824453\\
49	-1.00199928824453\\
50	-1.00260911867727\\
};
\addplot [color=mycolor4, forget plot]
  table[row sep=crcr]{%
1	0\\
2	0\\
3	-1.4579227673755\\
4	-1.4579227673755\\
5	-1.4579227673755\\
6	-0.905540105778007\\
7	-0.905540105778007\\
8	-0.905540105778007\\
9	-0.999388350959337\\
10	-0.999388350959337\\
11	-0.999388350959337\\
12	-1.0074894335959\\
13	-1.0074894335959\\
14	-1.0074894335959\\
15	-1.03275619388531\\
16	-1.03275619388531\\
17	-1.03275619388531\\
18	-1.00823380205307\\
19	-1.00823380205307\\
20	-1.00823380205307\\
21	-1.00532063934751\\
22	-1.00532063934751\\
23	-1.00532063934751\\
24	-1.0078420341821\\
25	-1.0078420341821\\
26	-1.0078420341821\\
27	-1.00515364025596\\
28	-1.00515364025596\\
29	-1.00515364025596\\
30	-1.00290742042252\\
31	-1.00290742042252\\
32	-1.00290742042252\\
33	-1.00158180677856\\
34	-1.00158180677856\\
35	-1.00158180677856\\
36	-1.00047251448741\\
37	-1.00047251448741\\
38	-1.00047251448741\\
39	-0.999914230917143\\
40	-0.999914230917143\\
41	-0.999914230917143\\
42	-1.00033474263539\\
43	-1.00033474263539\\
44	-1.00033474263539\\
45	-1.00176934485329\\
46	-1.00176934485329\\
47	-1.00176934485329\\
48	-0.998370912425659\\
49	-0.998370912425659\\
50	-0.998370912425659\\
};
\end{axis}
\end{tikzpicture}%

%% file: code2/temp_topo.tikz
\definecolor{mycolor1}{rgb}{0.00000,0.44700,0.74100}%
\definecolor{mycolor2}{rgb}{0.85000,0.32500,0.09800}%
\begin{tikzpicture}

\begin{axis}[%
width=1.566in,
height=1.566in,
at={(1.011in,0.642in)},
scale only axis,
xmin=0,
xmax=20,
ymin=0,
ymax=20,
axis background/.style={fill=white},
axis x line*=bottom,
axis y line*=left
]
\addplot [color=mycolor1, forget plot]
  table[row sep=crcr]{%
0	0\\
20	0\\
20	20\\
0	20\\
0	0\\
};
\addplot[only marks, mark=x, mark options={}, mark size=2.5000pt, draw=mycolor2] table[row sep=crcr]{%
x	y\\
9.34044009405148	8.3838902880659\\
15.4064898688432	13.7043900079352\\
1.00228749634689773	4.08904499463035\\
7.0466514526368	17.5623487278189\\
3.93511781634226	0.547751863958523\\
2.84677189537596	13.409350203568\\
4.72520422755342	8.34609604734254\\
7.91121454086096	11.173796568915\\
8.9353494846134	2.80773877190468\\
11.7763346800671	3.96202978169758\\
};
\end{axis}
\end{tikzpicture}%